%% file: paper.tex
\documentclass[12pt]{article}
\pdfoutput=1
\usepackage{jheppub}

\usepackage{amsmath,bbm,array,amsfonts,graphicx,wrapfig,lscape,float,slashbox,multirow,longtable,rotating,subfigure,epstopdf}

\input{pref}
\def\beqa{\begin{eqnarray}}
\def\eeqa{\end{eqnarray}}
\def\NN{{\cal N}}

\newcolumntype{C}[1]{>{\centering\arraybackslash}m{#1}}

\newcommand{\IS}{{\bf S}}
\newcommand{\IZ}{{\mathbb{Z}}}
\def\II{\relax{\rm I\kern-.18em I}}

\def\makeatletter{\catcode`\@=11}
\makeatletter
\def\mathbox#1{\hbox{$\m@th#1$}}%
\def\math@ccstyles#1#2#3#4#5#6#7{{\leavevmode
     \setbox0\mathbox{#6#7}%
     \setbox2\mathbox{#4#5}%
     \dimen@ #3%
     \baselineskip\z@\lineskiplimit#1\lineskip\z@
     \vbox{\ialign{##\crcr
            \hfil \kern #2\box2 \hfil\crcr
            \noalign{\kern\dimen@}%
            \hfil\box0\hfil\crcr}}}}
\def\mathaccstyles{\math@ccstyles\maxdimen}
\def\maththroughstyles{\math@ccstyles{-\maxdimen}}
\def\unity%
{\maththroughstyles{.45\ht0}\z@\displaystyle {\mathchar"006C}\displaystyle 1}

\newcommand{\drawsquare}[2]{\hbox{%
\rule{#2pt}{#1pt}\hskip-#2pt
\rule{#1pt}{#2pt}\hskip-#1pt
\rule[#1pt]{#1pt}{#2pt}}\rule[#1pt]{#2pt}{#2pt}\hskip-#2pt
\rule{#2pt}{#1pt}}

\newcommand{\fund}{~\raisebox{-.5pt}{\drawsquare{6.5}{0.4}}~}
\newcommand{\antifund}{~\overline{\raisebox{-.5pt}{\drawsquare{6.5}{0.4}}}~}

\newcommand{\symm}{~\raisebox{-.5pt}{\drawsquare{6.5}{0.4}}\hskip-0.4pt%
        \raisebox{-.5pt}{\drawsquare{6.5}{0.4}}~}

\newcommand{\asymm}{~\raisebox{-3.5pt}{\drawsquare{6.5}{0.4}}\hskip-6.9pt%
        \raisebox{3pt}{\drawsquare{6.5}{0.4}}~}

\newcommand{\antiasymm}{~\overline{\raisebox{-3.5pt}{\drawsquare{6.5}{0.4}}\hskip-6.9pt%
        \raisebox{3pt}{\drawsquare{6.5}{0.4}}}~}

\newcommand{\antisymm}{~\overline{\raisebox{-.5pt}{\drawsquare{6.5}{0.4}}\hskip-0.4pt%
        \raisebox{-.5pt}{\drawsquare{6.5}{0.4}}}~}

\title{D-brane Instantons as Gauge Instantons \\in Orientifolds of Chiral Quiver Theories}

\author[a,b]{Sebasti\'an Franco}
\author[c,d]{, Ander Retolaza}
\author[c]{, Angel Uranga}

\affiliation[a]{Physics Department, The City College of the CUNY \\ 160 Convent Avenue, New York, NY 10031, USA  }
\affiliation[b]{The Graduate School and University Center, The City University of New York \\ 365 Fifth Avenue, New York NY 10016, USA }
\affiliation[c]{Instituto de F\'isica Te\'orica IFT-UAM/CSIC \\
C/ Nicol\'as Cabrera 13-15, Universidad Aut\'onoma de Madrid,  28049 Madrid, Spain
}
\affiliation[d]{ Departamento de F\'isica Te\'orica, Universidad Aut\'onoma de Madrid,\\ Campus de Cantoblanco, 28049 Madrid, Spain }
\emailAdd{sfranco@ccny.cuny.edu, ander.retolaza@uam.es, angel.uranga@uam.es}

\abstract{Systems of D3-branes at orientifold singularities can receive non-perturbative D-brane instanton corrections, inducing field theory operators in the 4d effective theory. In certain {\em non-chiral} examples, these systems have been realized as the infrared endpoint of a Seiberg duality cascade, in which the D-brane instanton effects arise from strong gauge theory dynamics. We present the first UV duality cascade completion of {\em chiral} D3-brane theories, in which the D-brane instantons arise from gauge theory dynamics. Chiral examples are interesting because the instanton fermion zero mode sector is topologically protected, and therefore lead to more robust setups. As an application of our results, we provide a UV completion of certain D-brane orientifold systems recently claimed to produce conformal field theories with conformal invariance broken only by D-brane instantons.
}

\preprint{
\begin{flushright}CCNY-HEP-15-05 \end{flushright} \vspace{-0.9cm}
\begin{flushright}IFT-UAM/CSIC-15-073\end{flushright} \vspace{-0.9cm}
}

\begin{document}

\maketitle


\section{Introduction}

D-brane instantons can generate superpotential contributions to gauge theories on D-branes, which are forbidden to all orders in perturbation theory by global symmetries  \cite{Blumenhagen:2006xt,Ibanez:2006da,Florea:2006si}, see \cite{Blumenhagen:2009qh,Ibanez:2012zz} for reviews. These global symmetries, often anomalous, arise from the $U(1)$ factors inside the $U(N_i)$ groups associated to $N_i$ fractional branes of a given type. The contributions of D-brane instantons are exponentially sensitive to the volume of the cycles they wrap. There is a vast list of scenarios in which such effects are crucial. Just to mention a few, they can generate neutrino masses \cite{Blumenhagen:2006xt,Ibanez:2006da} (see also \cite{Ibanez:2007rs}), Yukawa couplings \cite{Blumenhagen:2007zk}, the $\mu$-term in SUSY extensions of the Standard Model \cite{Cvetic:2010dz,Cvetic:2010mm}, or be crucial in SUSY breaking \cite{ Argurio:2006ny,Argurio:2007qk,Cvetic:2008mh} or its mediation \cite{Buican:2008qe}, as well as in rare processes \cite{Blumenhagen:2010dt,Addazi:2014ila,Addazi:2015rwa,Addazi:2015hka} (see \cite{Blumenhagen:2009qh,Ibanez:2012zz} for reviews of these and other applications). Achieving these effects typically requires introducing orientifold projections to remove additional fermion zero modes associated to the otherwise underlying $\NN=2$ supersymmetry \cite{Argurio:2007vqa,Bianchi:2007wy,Ibanez:2007rs} and \cite{Blumenhagen:2009qh,Ibanez:2012zz} for reviews.

In some cases, when D-brane instantons sit on top of gauge D-branes, they can be interpreted as ordinary gauge instantons. In other more general situations, the D-brane instantons are a consequence of the UV completion of the theory. In other words, restricting the dynamics to the naive low-energy gauge theory one loses the information about the possible presence of these D-brane instantons. These two possibilities have led to distinction between `gauge' and `exotic' D-brane instantons. Interestingly, in some situations, it is also possible to derive exotic instantons from an alternative, fully gauge theoretic, UV completion \cite{Aharony:2007pr}, showing that the distinction is to some extent an artifact of the low-energy truncation. This completion involves a cascade of Seiberg dualities and some non-trivial IR dynamics of the gauge theories. 

By now, there are various explicit examples in the literature for which this has been achieved \cite{Aharony:2007pr,Argurio:2012iw}.\footnote{See also \cite{GarciaEtxebarria:2008iw,Krefl:2008gs} for similar conclusion in different approaches.} However, all these examples correspond to theories that, before orientifolding, are non-chiral. This may be problematic in certain applications when these theories are combined with extra ingredients, which may remove the non-chiral instanton fermion zero modes, and consequently the insertions of charge matter fields in the instanton amplitude evaporate. An interesting analysis of the last Seiberg duality transformations, but not the full cascades, leading to D-brane instantons in some chiral theories has appeared in \cite{Amariti:2008xu}.

One of the main goals of this article is to construct similar UV completions for chiral theories.\footnote{For brevity, in what follows we refer as chiral to theories that are indeed chiral even before orientifolding or adding flavors.} This is can be taken as a mere proof of existence, but can be relevant for the applications mentioned above, in which now the chiral nature of the theory protects the instanton fermion zero modes, and the presence of charged matter fields in the instanton induced operator is also protected. Another novel feature of some of the theories we consider, absent from previous constructions in the literature, is the presence of D-brane instanton couplings involving flavors. 

As an application of the ideas introduced in this paper, we provide a UV duality cascade completion and gauge theory derivation for the exotic D-brane instantons that break conformal invariance in the models considered in \cite{Bianchi:2013gka}.  

This paper is organized as follows. \sref{section_intro_instantons} contains a lightning review of D-brane instantons and the field theory couplings they can generate. \sref{section_field_theory} summarizes the general approach used for finding cascading UV completions of D-brane instantons. In \sref{sec:cascading} we describe chiral IR theories that can be directly completed in terms of cascades, which maintain the underlying geometry at which D3-branes are located. We present an explicit example of this kind based on an orientifold of a ${\mathbb{Z}}_2$ orbifold of the conifold (the $\mathbb{F}_0$ theory). In \sref{sec:non-cascading} we consider chiral IR theories that do not correspond directly to cascading geometries, but can be UV completed in terms of a duality cascade which, upon complex deformation (confinement in a subset of their nodes) produced the IR theory of interest. We present an explicit example based on an orientifold of a ${\mathbb{Z}}_3$ orbifold of the SPP singularity deformed to an orientifold of $\mathbb{C}^3/\mathbb{Z}_3$. Since the earlier examples correspond to orientifolds by line reflections in the dimer diagram representation of the field theory, \sref{sec:opoints} briefly discusses models based on orientifold actions corresponding to point reflection in the dimer diagram. In \sref{section_CFT_breaking} we discuss the introduction of flavor D7-branes under which some of the instanton fermion zero modes are charged, and provide an illustrative example providing a UV completion of the models argued in \cite{Bianchi:2013gka} to have conformal invariance only broken by D-brane instanton effects.  We conclude and discuss future directions in \sref{section_conclusions}. Appendix \ref{sec:thespp} describes some aspects of the SPP theory and its orientifolds.

\bigskip

\section{Superpotential Couplings from D-brane Instantons}

\label{section_intro_instantons}

In this section we will quickly review the basic ideas on how gauge theories on stacks of space-time filling D-branes can be perturbed by superpotential couplings generated by Euclidean Dp-brane instantons \cite{Blumenhagen:2006xt,Ibanez:2006da,Florea:2006si}.\footnote{D-brane instantons can lead to other modifications of these gauge theories, by higher derivative operators, but this is beyond the focus of this article. For some examples, see e.g. \cite{GarciaEtxebarria:2007zv}.} More detailed techniques will be explained in later sections as needed.

\begin{figure}[!ht]
\begin{center}
\includegraphics[width=11cm]{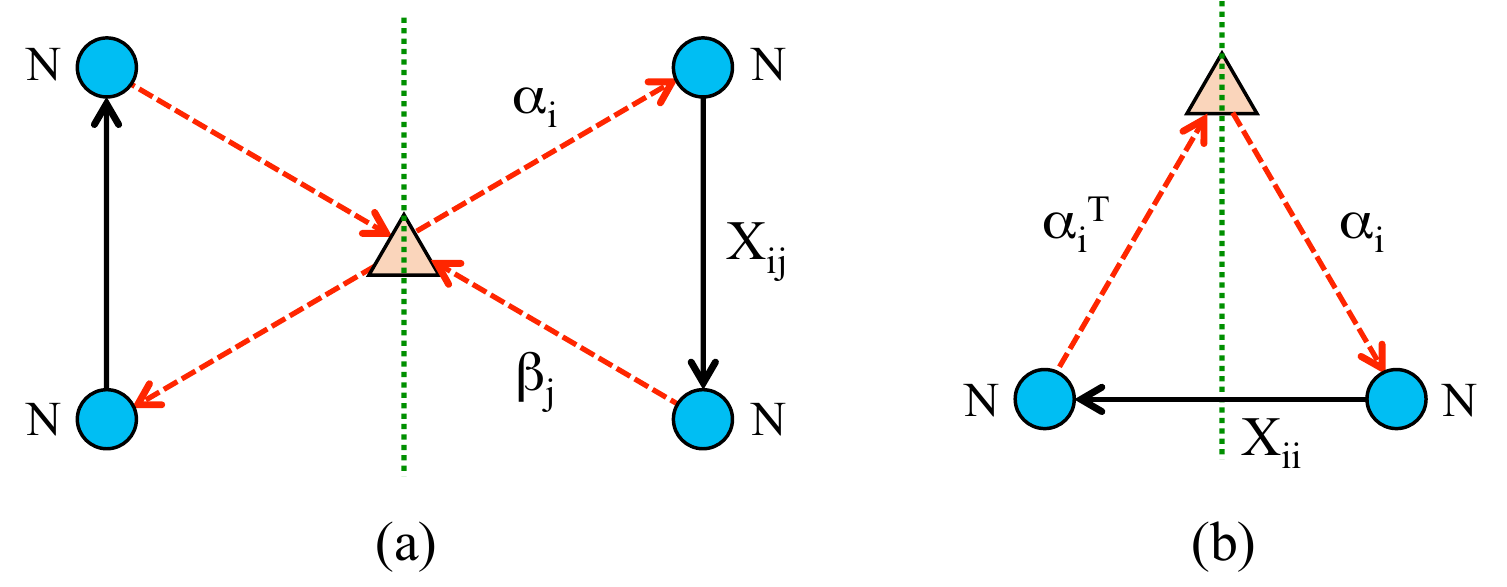}
\caption{(a) Quiver for an instanton with zero modes coupling to a field not invariant under the orientifold action. The green line represents the fixed locus under the orientifold action. No D-brane instanton coupling is generated if the ranks of the nodes connected by $X_{ij}$ differ. (b) Quiver for an instanton with zero modes coupling to a field mapped to itself under the orientifold action.}
\label{fig:zeromodes}
\end{center}
\end{figure}

Let us first consider the extended quiver diagram in \fref{fig:zeromodes}(a), which encodes the relevant field content for a configuration with D-brane instantons and 4d spacetime filling gauge D-branes. In this figure, circles correspond to two $SU(N)$ gauge groups in the quiver living on 4d space-filling branes which, in principle, might contain additional nodes and chiral fields. The ranks of both nodes need to be equal for a non-vanishing instanton contribution to exist.\footnote{Clearly, there may be situations where the instanton may have incoming and/or outgoing arrows from/to more nodes, in which case their ranks can be unequal, only the total numbers should match.} $X_{ij}$ can correspond to a single bifundamental field or, more generally, to a product of them of the form $X_{ij}=X_{i k_1} X_{k_1 k_2}\ldots X_{k_n j}$, where intermediate color indices are summed. The explicit form of $X_{ij}$ is controlled by the geometry of the D-brane instanton under consideration. We refer the reader to \cite{Blumenhagen:2009qh,Ibanez:2012zz} for thorough discussions of this point. The D-brane instanton is represented by the triangular node in \fref{fig:zeromodes}. There are charged fermionic zero modes $\alpha_i$ and $\beta_j$ at the intersections between the instanton and the gauge D-branes. The instanton action contains the following term
\beq
\mathcal{L}=\alpha_i X_{ij} \beta_j .
\eeq

Strings with both endpoints on the instanton give rise to neutral zero modes. Instantons break 1/2 of the $\mathcal{N}=1$ SUSY preserved by the D-branes at the Calabi-Yau and hence there are two fermionic zero modes, the corresponding goldstinos, which are represented by Grassmann variables $\theta^\alpha$. However, the sector of open strings with both endpoints on the instanton D-brane sees an enhanced $\mathcal{N}=2$ SUSY so we should generically expect two additional fermionic zero modes. In order to generate non-vanishing superpotential couplings, it is necessary to have only two fermionic zero modes, which are used to saturate the superspace measure. It is possible to imagine various ways in which the two extra zero modes can be in principle eliminated. A particularly clean way of getting rid of them, which will be realized in the theories we consider, is by projecting them out with an orientifold plane reducing the instanton worldvolume group to $O(1)$ (these are dubbed $O(1)$ instantons). We are thus lead to consider configurations in which the D-brane instanton is on top of an orientifold plane, \fref{fig:zeromodes}.a. Consider first the case where the orientifold is not relating the two $SU(N)$ nodes under which fermion zero modes are charged (i.e. the operator $X_{ij}$ is not mapped to itself under the orientifold action). Integrating out the zero modes, the following contribution to the gauge theory superpotential is generated
\beq
W_{inst}=M_s^{3-N} e^{-V_\Sigma/g_s} \det X_{ij}\ ,
\label{W_inst_1}
\eeq
with $M_s$ the string scale and $V_\Sigma$ the volume of the cycle $\Sigma$ wrapped by the instanton in string units and have assumed a numerical constant in front of the expression is $\mathcal{O}(1)$. 

Consider now the case that the orientifold identifies the two $SU(N)$ nodes of the quiver, \fref{fig:zeromodes}(b) For $O(1)$ instantons, the orientifold we are interested in is such that the operator $X_{ij}$ is projected down to the antisymmetric representation (or its conjugate) of the surviving $SU(N)$. The orientifold also identifies the fermions zero modes $\beta_j\sim \alpha_j{}^T$, so the instanton action now contains an interaction of the form
\beq
\mathcal{L}=\alpha_i X_{ii} \alpha_i{}^T \, .
\eeq
After integration over all fermionic zero modes, the superpotential picks a contribution of the following form
\beq
W_{inst}=M_s^{3-N/2} e^{-V_\Sigma/g_s} {\rm Pf} \, X_{ij}\ .
\label{W_inst_2}
\eeq
It is important to emphasize an important difference between non-chiral and chiral quivers. Non-chiral quivers can be regarded as a subset of the theories described above for which both $SU(N)$ nodes happen to be the same, whereas chiral examples correspond to the generic case. This has an impact on the behavior of the instantons with respect to $U(1)$ symmetries. Actually, the gauge groups on the gauge D-branes are $U(N)=SU(N)\times U(1)$. The $U(1)$ factors either become massive due to the Green-Schwarz anomaly cancellation mechanism \cite{Ibanez:1998qp} or become free in the IR. For non-chiral theories, the operator $X_{ij}$ is uncharged under the $U(1)$'s, and so is the field theory operator induced by the instanton. On the other hand, chiral examples produce field theory operators violating the $U(1)$ symmetries. It is therefore in the chiral examples that $U(1)$ symmetries forbid couplings such as \eref{W_inst_1} and \eref{W_inst_2} perturbatively, and are generated by the instantons. The fact that the instanton amplitude implies some charge violation is related to the chiral nature of the charged fermion zero modes, and is robust under deformations of the theory.

\bigskip

\section{General Approach to Field Theoretic UV Completions}

\label{section_field_theory}

Before presenting a detailed analysis of a cascading UV completion of the theories of interest, it is useful to present a general roadmap summarizing its key points. While this class of theories has novel features which will be emphasized as they appear, the general approach is similar to the one used in other examples \cite{Aharony:2007pr,Amariti:2008xu,Argurio:2012iw}. Some readers might also find this outline handy for applications to other theories.

The starting point is a gauge theory, which we regard as the IR of a more complete UV configuration that includes extra ingredients to support the instanton generating the non-perturbative correction.  In principle, it is possible to take a bottom-up approach and try to guess a UV quantum field theory that results in the desired one at low energies. This is however rather challenging, since some of the gauge groups may disappear while flowing to the IR, by confinement or other strong coupling phenomena.
Our strategy is based in instead using string theory constructions to determine the UV theory. Namely, we describe string theory configurations of D-branes at singularities to engineer the IR theory, and use well-established tools inspired by holography to construct UV completions with the appropriate IR flow. We anticipate that we will focus on the very tractable setup of D3-branes at toric CY threefold singularities, which have been extensively studied using brane tilings (a.k.a. dimer diagrams) \cite{Franco:2005rj,Franco:2005sm} (see also \cite{Kennaway:2007tq,Yamazaki:2008bt} and references therein).\footnote{For generalizations known as Bipartite Field Theories, see \cite{Franco:2012mm,Xie:2012mr,Franco:2012wv,Heckman:2012jh,Franco:2013pg,Franco:2013ana,Franco:2013nwa,Franco:2014nca}.
In addition, these theories have also appeared recently in the context of class $S_k$ theories in \cite{Gaiotto:2015usa,Franco:2015jna,Hanany:2015pfa}.} Their orientifold quotients can be constructed systematically using the tools in \cite{Franco:2007ii}.

The general strategy we follow to construct such UV completions can be summarized as follows:

\begin{itemize}
\item   The first step is to engineer the geometry probed by the D-branes realizing the IR gauge theory. This typically corresponds to a quiver gauge theory, which must involve additional empty nodes to support the stringy instantons. As explained in \sref{section_intro_instantons}, the instantons need to have an $O(1)$ orientifold projection, hence the opposite orientifold action on spacefilling D-branes leads to $USp$ gauge groups. For the cases considered in this article, the gauge group takes the form $\prod_i SU(N_i) \times \prod_a USp(N_a)$.\footnote{Even restricting to D-branes at toric singularities, it is possible to have multiple $USp$ nodes, and also to simultaneously include $SO$ ones. Since the latter do not support interesting instantons in the IR, for notational simplicity we ignore the $SO$ factors in the discussions.}

\item The next step is to construct the corresponding duality cascade. In principle, it is necessary to specify the ranks and dynamical scales of the gauge groups (even for those empty in the IR theory) at some UV scale, since this information determines the sequence of dualizations. In practice, we will exploit examples where this analysis has already been carried out (or orbifolds thereof), and apply the dualities with the rule of thumb that the node that is dualized at each step is asymptotically free.
 Cascades are typically periodic, i.e. all nodes in the quiver are dualized in the cascade and produce, after a `period' involving a number of dualities, a theory of the same form as the original one, up to an overall decrease in the ranks of all the gauge groups. If the IR theory does not admit a direct UV completion with a duality cascade (we dub them `\textit{non-cascading}' systems), one must embed them as the IR result of a confinement process of a more involved theory with additional nodes, such that the latter does admit a cascade completion in its UV. We will find explicit examples of both kinds of behaviors.

\item
In some of our models, we will be interested in introducing additional flavors, obtained by enriching the systems of D3-branes at singularities with additional D7-branes. These extra flavors modify the strict periodicity of the cascades, by ${\cal O}(1/N_i)$ corrections, where $N_i$ denotes the ranks of the gauge factors. In the gravity dual counterpart, they correspond to numbers of D3-branes, and the corrections are due to one-loop effects from the D7-brane backreaction; ignoring them corresponds to the familiar probe approximation, in which the D7-branes simply probe the geometry generated by the fractional D3-brane duality cascade and their IR deformation.

\item The last step is to verify that, upon running the cascade down, one recovers the original IR gauge theory of interest (modified by the non-perturbative operator). This heuristically corresponds to the extrapolation of the cascade until the number of D3-branes reaches a lower bound. More rigorously, the procedure takes the theories beyond the range of validity of Seiberg duality; so we must instead analyze the strong dynamics of the last steps in the cascade. This sometimes involves confinement and a complex deformation of the moduli space, reflecting geometric transitions in the underlying singularities.

\end{itemize}

All these ideas are illustrated in detail in the explicit examples considered in the next sections.

\bigskip

\section{Cascading Geometries}
\label{sec:cascading}

\subsection{Cascading Versus Non-Cascading Geometries}

It is convenient to classify IR theories according to whether they admit fractional branes, which trigger duality cascades, or not. We refer to the two resulting classes as {\it cascading} and {\it non-cascading} geometries, respectively. All examples in the literature in which D-brane instanton couplings have been UV completed by cascading geometries associated to non-chiral theories. Since fractional branes are related to anomaly free rank assignments for gauge groups, non-cascading geometries can only arise when considering chiral theories. The UV completion of non-cascading geometries requires additional ingredients, which will be presented in \sref{sec:non-cascading}. In this section, we consider the simpler case of chiral theories for cascading geometries.

\bigskip

\subsection{D-brane Instanton Couplings}

\begin{figure}[!ht]
\begin{center}
\includegraphics[width=4cm]{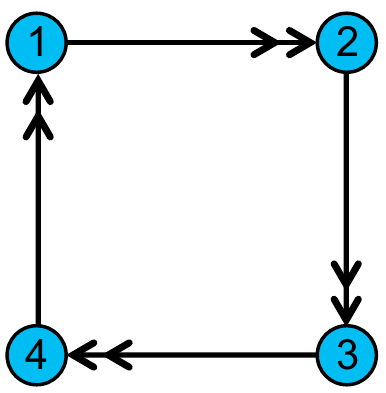}
\caption{Quiver diagram for $\mathbb{F}_0$.}
\label{quiver_F0}
\end{center}
\end{figure}

For concreteness we carry out the discussion for a prototypical example, the $\mathbb{F}_0$ theory, which corresponds to a chiral $\mathbb{Z}_2$ orbifold of the conifold. The quiver diagram for one of the two toric phases of $\mathbb{F}_0$ is given in \fref{quiver_F0}, where the ranks $N_i$ of the gauge groups  correspond to the numbers of (fractional) branes wrapped at the singularity and are constrained by the cancellation of non-abelian anomalies.\footnote{The $\mathbb{F}_0$ theory has another toric phase \cite{Feng:2002zw}, related to this gauge theory by Seiberg duality \cite{Feng:2001bn}; it will appear in intermediate steps of the duality cascade in \sref{section_cascade_F0}.} The anomaly constraints are modified in the presence of flavor branes, which will be introduced later on, so for the moment we keep the ranks general and focus on the structure of the quiver theory. The superpotential is 
\beq
W=X^1_{12}X^1_{23}X^2_{34}X^2_{41}-X^1_{12}X^2_{23}X^2_{34}X^1_{41}-X^2_{12}X^1_{23}X^1_{34}X^2_{41}+X^2_{12}X^2_{23}X^1_{34}X^1_{41}\, .
\label{W_F0}
\eeq
Here and throughout the article, we leave an overall trace over color indices in the superpotential implicit. The theory has an $SU(2)\times SU(2)$ global symmetry under which the fields on each side of the square transform as doublets, whose components are indicated by the superindices.

\begin{figure}[!ht]
\begin{center}
\includegraphics[width=11cm]{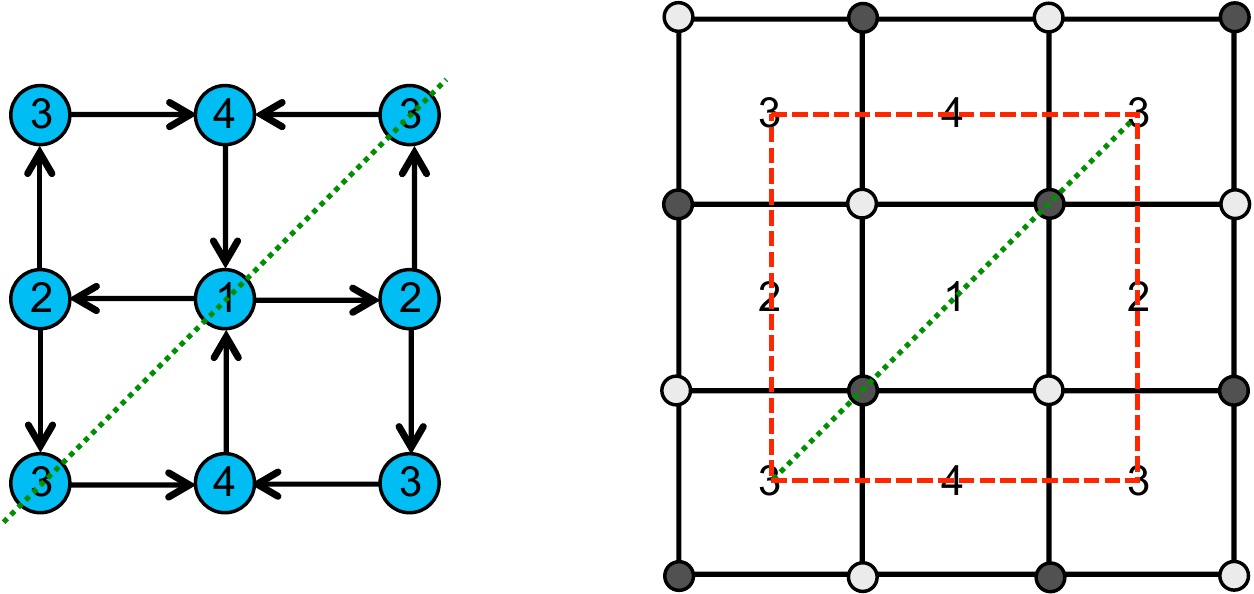}
\caption{The periodic quiver and dimer diagram for the $\mathbb{F}_0$ theory.}
\label{periodic_quiver_dimer_F0}
\end{center}
\end{figure}

In order to properly analyze orientifolds of toric CY 3-folds, it is convenient to consider brane tilings, and their dual periodic quivers \cite{Franco:2007ii}. We show them for this theory in \fref{periodic_quiver_dimer_F0}. The orientifold corresponds to a $\IZ_2$ symmetry of the periodic quiver, flipping the orientation of the arrows. In the following we consider the orientifold associated to a reflection with respect to the line shown in \fref{periodic_quiver_dimer_F0}. This theory also admits orientifolds associated to fixed points in the dimer, but we will not consider this possibility (see \sref{sec:opoints} for orientifolds with fixed points in the dimer).

We choose the orientifold line charge such that the instanton has a worldvolume $O(1)$ group to get rid of unwanted fermion zero modes; then the action on gauge branes is of $USp$ kind. The orientifolded theory has three gauge groups \footnote{We take the convention in which $USp(N_c)=Sp(N_c/2)$, where $N_c$ is an even number. This convention has the advantage of capturing the number of branes in the configuration in a simple way: $USp(N_c)$ gauge group arises from $N_c$ D-branes in the parent theory at the corresponding node, which is fixed under the orientifold. The number of flavors $N_f$ is defined as the number of number of chiral multiplets in the fundamental representation (equivalently, half the chiral multiplets in the fundamental representation in the covering space), which is $N_c$-dimensional.\label{sp-convention}}
\beqa
USp(N_1) \times USp(N_3) \times U(N_{2})
\eeqa
and four bifundamental chiral multiplets, since the orientifold identifies nodes $2\leftrightarrow 4$ and relates $X^a_{41}\leftrightarrow X^a_{12}{}^T$ and  $X^a_{34} \leftrightarrow X^a_{23}{}^T$ for $a=1,2$. We choose the convention of preserving fields (and coupling) on the right hand side of the orientifold line in the dimer. The corresponding superpotential can be obtained by truncating \eref{W_F0} onto invariant states, and becomes
\beqa
W=X^1_{12}X^1_{23}X^2_{23}{}^TX^2_{12}{}^T-X^1_{12}X^2_{23}X^2_{23}{}^TX^1_{12}{}^T-X^2_{12}X^1_{23}X^1_{23}{}^TX^2_{12}{}^T\, ,
\label{W_classic_F0_orientifolded}
\eeqa
with appropriate contractions of the color indices.\footnote{Since we have $SU$ and $USp$ groups, one must contract each kind of index in a different way. $SU$ indices are directly contracted, while $USp$ ones are contracted with the $USp$ invariant tensor $J$.} Two of the original terms are identified, resulting in three terms. The transposition arises when relating fields to their orientifold images and allows the standard contraction of color indices. This is in agreement with the structure of the superpotential in the orientifolded dimer, shown in \fref{periodic_quiver_dimer_F0}, as two of the nodes in the dimer are identified by the orientifold. 

\begin{figure}[!ht]
\begin{center}
\includegraphics[width=6.5cm]{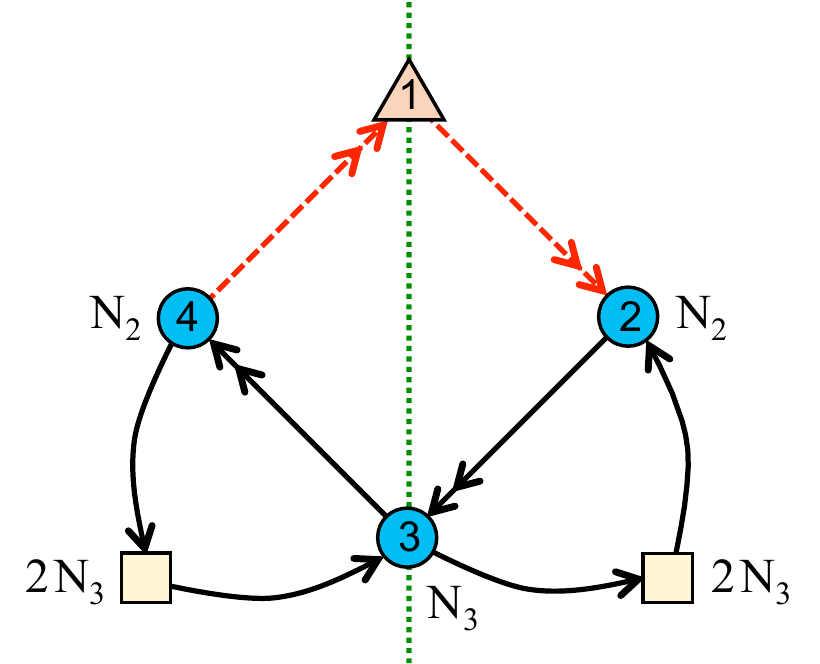}
\caption{The $\mathbb{F}_0$ quiver with an empty node, supporting a D-brane instanton. Here $N_3$ is an even number since the group 3 is of $USp$ kind.} 
\label{quiver_F0_instanton}
\end{center}
\end{figure}

We are interested in configurations such as the one shown in \fref{quiver_F0_instanton}, where the vertical green dashed line schematically represents the effect of the orientifold projection in the quiver and we added two stacks of flavor branes represented by square nodes (actually, one stack and its orientifold image). 
Node 1 is taken to be empty in order to support a `exotic' D-brane instanton. With our above choice of orientifold charge, the instanton has a worldvolume $O(1)$ group and the orientifold removes extra neutral fermion zero modes.
Since the action on gauge branes is of $USp$ kind, we must introduce at least two D3-branes at node 3. We include $N_2$ deformation fractional branes, which are responsible for driving the duality cascade in the UV, and additional ${\cal O}(1)$ D3- and D7-branes to render the structure anomaly free.\footnote{Throughout this article we adopt the classification of fractional branes according to the IR dynamics they generate introduced in \cite{Franco:2005zu}. In this classification, fractional branes are divided into three types: i) deformation branes, which trigger complex deformations in the supergravity dual, ii) $\NN=2$ branes, which lead to $\NN=2$ dynamics, and iii) DSB branes, which produce dynamical supersymmetry breaking.} We keep the ranks of nodes 2 and 3 and the number of D7-branes as general as possible, while consistent with anomaly cancellation. The D7-branes introduce D7-D3 flavors, with cubic couplings 73-33-37 to the D3-D3 chiral multiplets, described by triangles of arrows in the quiver. We will not need the detailed structure of these couplings since the purpose of the D7-D3 flavors is to cancel anomalies and play no further role in our analysis.\footnote{In addition, generically, mesons that are bifundamental of a pair of D7-brane nodes can be generated whenever a gauge group is dualized in a flavored quiver. Depending on the theory, such mesons can become massive after a number of steps in the cascade or accumulate. These chiral fields are neutral under all gauge symmetries and do not affect our analysis, so we will not include them in our discussion of any of the models in this article.}

The structure of zero modes and their couplings to D3-D3 fields are identical to those on space-filling D-branes. Hence, from \eref{W_F0} we can infer the following couplings in the instanton partition function of the unorientifolded theory
\beq
\lambda^1_{12}X^1_{23}X^2_{34}\lambda^2_{41}-\lambda^1_{12}X^2_{23}X^2_{34}\lambda^1_{41}-\lambda^2_{12}X^1_{23}X^1_{34}\lambda^2_{41}+\lambda^2_{12}X^2_{23}X^1_{34}\lambda^1_{41}\, .
\label{fermion_couplings_unorientifolded_F0}
\eeq
In addition to the D3-D3 field identifications, the orientifold relates $\lambda^a_{12} \leftrightarrow \lambda^a_{41}{}^T$, for $a=1,2$. Then, \eref{fermion_couplings_unorientifolded_F0} becomes
\beqa
\lambda^1_{12}X^1_{23}X^2_{23}{}^T \lambda^2_{12}{}^T - 
\lambda^1_{12}X^2_{23}X^2_{23}{}^T\lambda^1_{12}{}^T-
\lambda^2_{12}X^1_{23}X^1_{23}{}^T\lambda^2_{12}{}^T\, ,
\label{fzm-supo}
\eeqa
which can equivalently be obtained from \eref{W_classic_F0_orientifolded} by replacing some fields by zero modes.

Integrating out the charged fermionic zero modes, we obtain the following non-perturbative D-brane instanton superpotential:
\beqa
W \sim{\rm Pf} \, {\cal N}_{22} \, ,
\label{pf-one}
\eeqa
where ${\cal N}_{22}$ is the matrix of D3-D3 mesons of node 3 (note that it does not include the D3-D7 flavors)     
\be
{\cal N}_{22}=\left(\begin{array}{cc} {\cal N}^{11}_{22} \ & {\cal N}^{12}_{22}\ \\[.1 cm] {\cal N}^{21}_{22} & {\cal N}^{22}_{22}\end{array} \right)=\left(\begin{array}{cc} X^1_{23} X^1_{23}{}^T \ & X^1_{23} X^2_{23}{}^T \ \\[.1 cm] X^2_{23} X^1_{23}{}^T & X^2_{23} X^2_{23}{}^T \end{array} \right) \, .
\label{cal_N22_F0}
\eeq
In constructing $\mathcal{N}_{22}$ we should take into account the appropriate contractions of color indices at node 3 we alluded to before. The subindices emphasize that the resulting $\mathcal{N}_{22}$ transforms in the adjoint representation of node 2.

Before concluding, let us emphasize that the origin of (\ref{pf-one}) is the stringy instanton sitting on node 1, and not the strong dynamics of node 3.

\bigskip

\subsection{The Cascade}

\label{section_cascade_F0}

This theory has a periodic cascade of Seiberg dualities, which we now explain.

\bigskip

\subsection*{The Parent Cascade}

It is convenient to discuss the cascade in the unorientifolded theory first. This cascade has been investigated in detail in \cite{Franco:2003ja,Franco:2004jz,Franco:2005fd}, so our presentation will be brief. 

\begin{figure}[!ht]
\begin{center}
\includegraphics[width=11cm]{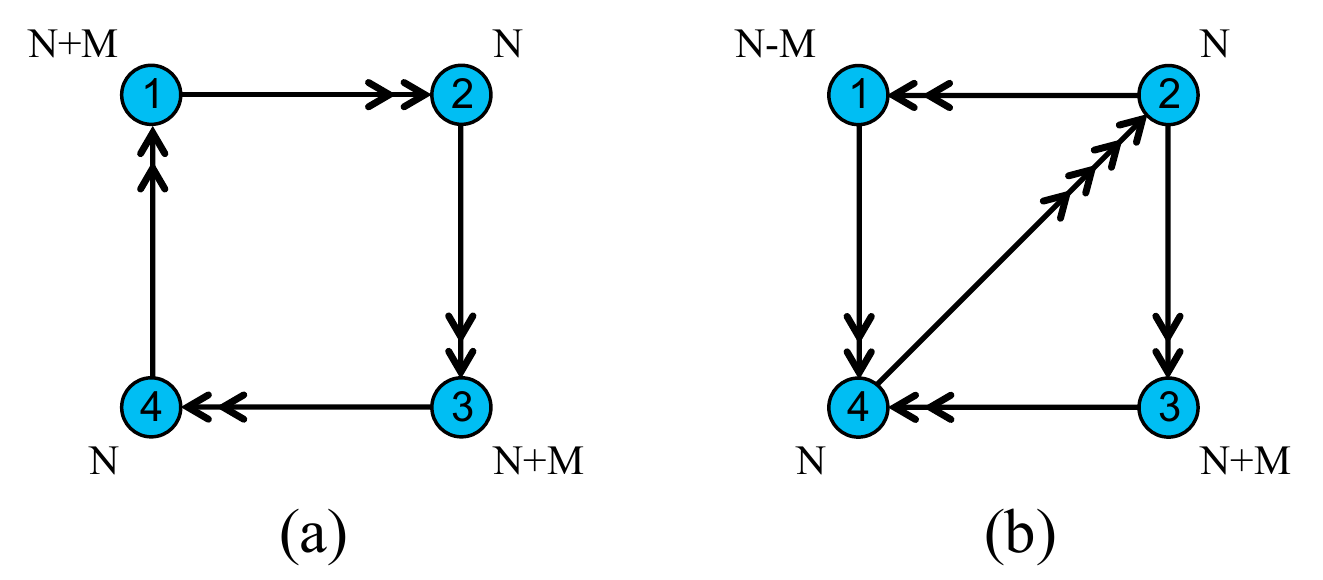}
\caption{(a) Quiver diagram for one of the phases of $\mathbb{F}_0$ for $N$ regular D3-branes and $M$ fractional branes. (b) The quiver diagram after Seiberg dualizing node 1.} 
\label{2_phases_F0}
\end{center}
\end{figure}

Consider starting from the quiver in \fref{2_phases_F0}(a), which corresponds to $N$ regular D3-branes and $M$ fractional branes, at some point in the UV. The periodic cascade corresponds to repeating the sequence of dualizations (1,3,2,4), i.e. sequentially dualizing pairs of diagonally opposite nodes. 

Let us consider the first step. Starting from \fref{2_phases_F0}(a) and dualizing node 1, we obtain the quiver shown in \fref{2_phases_F0}(b). The four chiral fields in the diagonal connecting nodes 4 and 2 arise as mesons of node 1. Dualizing node 3, we generate four new mesons stretching between nodes 2 and 4, in the opposite orientation. The superpotential contains mass terms for all fields in the diagonal, which disappear after integrating them out. After these two dualizations, we thus recover the original theory, up to a reversal of the direction of all arrows and a reduction in the ranks of nodes 1 and 3. It is hence clear that completing a period in the cascade by further dualizing nodes 2 and 4 brings us back to the original theory, where the number of D3-branes is reduced $N\to N-2M$ and the number of fractional branes $M$ remains constant. The effective number of D3-branes decreases logarithmically as a function of energy along the cascading RG flow. The cascade is just a $\IZ_2$ orbifold version of the conifold cascade in \cite{Klebanov:2000hb}.

\bigskip

\subsection*{The Orientifolded Cascade}

\begin{figure}[!ht]
\begin{center}
\includegraphics[width=13.5cm]{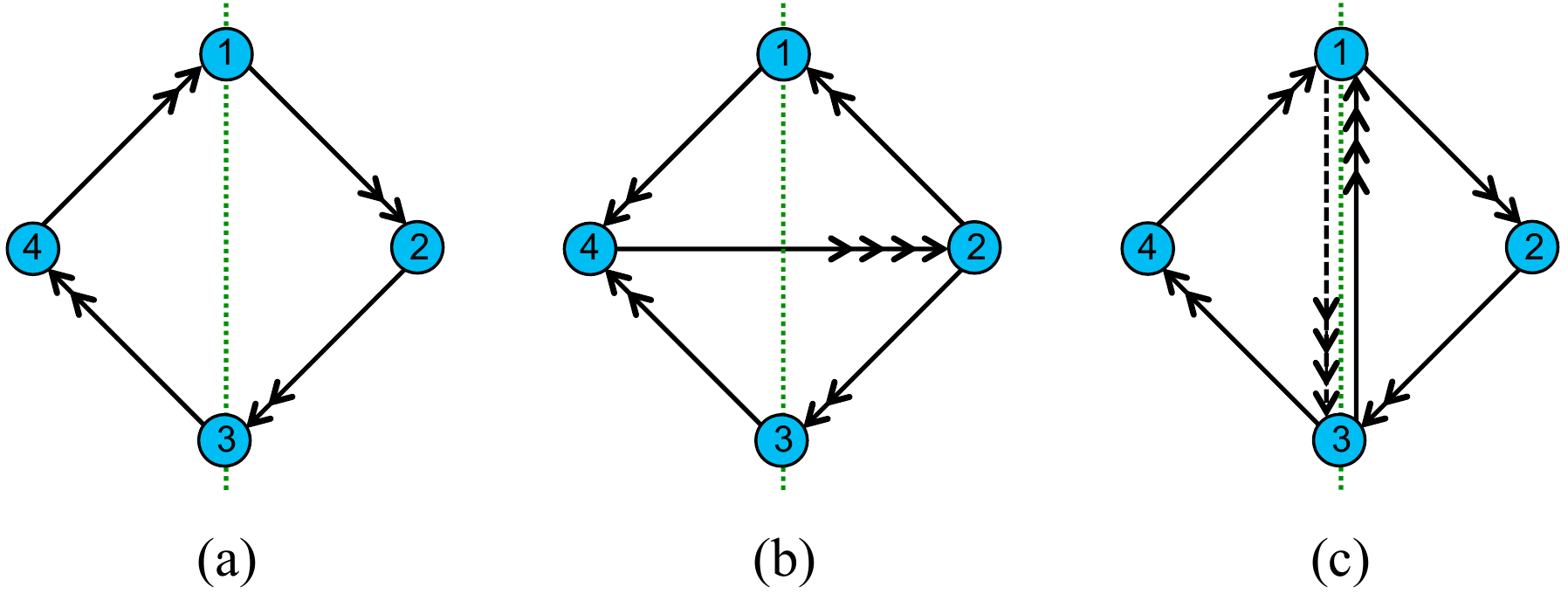}
\caption{The orientifolded cascade alternates between these three quivers. As explained below, the theory (c) is actually equivalent to (a).} 
\label{3_phases_F0_orientifolded}
\end{center}
\end{figure}

The orientifolded cascade starts from the quiver in \fref{3_phases_F0_orientifolded}(a) and repeats the sequence of dualizations (1,3,2), alternating between the quivers shown in \fref{3_phases_F0_orientifolded}.\footnote{More precisely, at some steps in the cascade we obtain quivers that are identical to those in \fref{3_phases_F0_orientifolded} up to an overall flip in the orientation of arrows. Also, note that 4 is the image of 2, so we can think both are dualized simultaneously.} In this quivers, arrows are oriented to keep track of the representation under the $U(N)$ gauge factor from nodes 2, 4 (but note that representations of the $USp$ groups at nodes 1, 3, are real, so the orientation {\em at} those nodes is meaningless). 

In general lines, the dualizations of nodes 1 and 3 work as in the parent theory, although using the $USp$ version of Seiberg duality \cite{Seiberg:1994pq}.  Let us first recall how Seiberg duality works for a $USp(N_c)$ gauge group with $N_f$ flavors, namely $N_f$ chiral fields in the fundamental representation (c.f. footnote \ref{sp-convention} for conventions). The dual theory has $USp(N_f-N_c-4)$ gauge group, $N_f$ flavors, and mesons with cubic superpotential couplings to the flavors. We refer the reader to \cite{Intriligator:1995ne} for additional details. In our present setup, when we dualize node 1, there appear mesons connecting nodes 4 and 2, as shown in \fref{3_phases_F0_orientifolded}(b). These mesons transform in the conjugate antisymmetric representation of node 2. They become massive by combining with the oppositely oriented mesons that appear upon dualization of node 3. The result is a diagram similar to \fref{3_phases_F0_orientifolded}(a), up to reversal of arrows.

When we subsequently dualize node 2, we obtain the quiver shown in \fref{3_phases_F0_orientifolded}(c). This step gives rise to mesons connecting nodes 1 and 3. In terms of the parent quiver, this step actually also involves the dualization of its orientifold image, node 4, which would give rise to the dashed fields in \fref{3_phases_F0_orientifolded}(c). Moreover, the superpotential \eref{W_classic_F0_orientifolded} gives rise to masses for all the mesons and the theory becomes the one in \fref{3_phases_F0_orientifolded}(a) after integrating them out. We conclude that the sequence of dualizations (1,3,2) corresponds to a period in the cascade.

\bigskip

\subsection{The IR Bottom of the Cascade}
\label{sec:irbottom}

We now explain how \eref{pf-one} can be alternatively understood as resulting from the gauge dynamics at the IR bottom of the duality cascade discussed in \sref{section_cascade_F0}, in which the instanton is realized in terms of a standard gauge instanton. To do so, let us consider the theory a couple of steps before the one in \fref{quiver_F0_instanton}. We can quickly determine the corresponding quiver by Seiberg dualizing nodes 3 and 1. The latter is just a formal dualization of an empty node, but it will acquire more physical significance when we consider how the strong dynamics of the UV gauge theory indeed reproduces the properties of the empty node, as shown in \sref{section_useful_trick}.

\begin{figure}[!ht]
\begin{center}
\includegraphics[width=6.5cm]{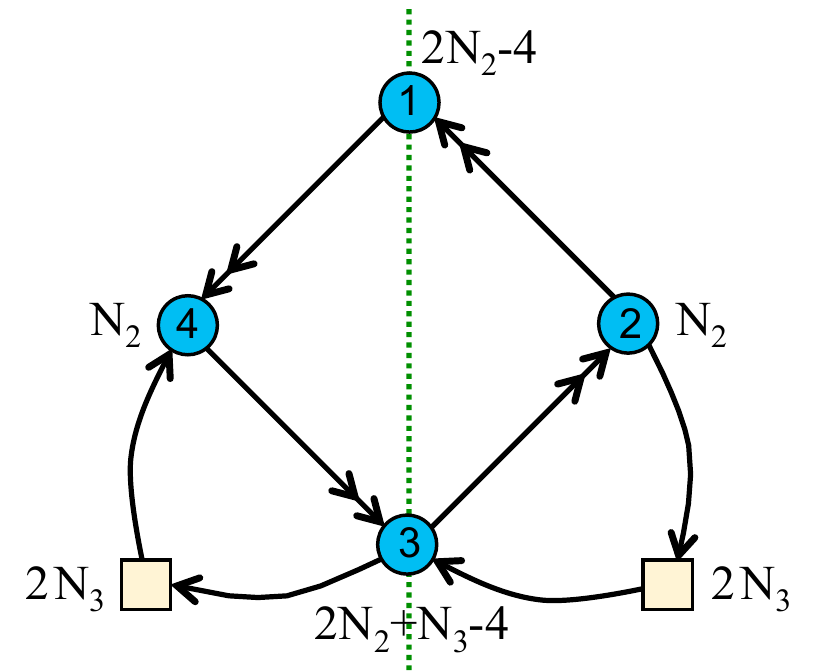}
\caption{The quiver of the orientifolded $\mathbb{F}_0$ theory, two steps before the IR bottom of the cascade.} 
\label{quiver_F0_last_step}
\end{center}
\end{figure}

This formal Seiberg duality on nodes 1 and 3 in \fref{quiver_F0_instanton} produces the quiver shown in \fref{quiver_F0_last_step}. All arrows connected to the dualized nodes have been inverted. We note that the dualization of node 3 produces several mesons. First, there are mesons connecting node 2 to its orientifold image, which become massive with mesons arising upon dualization of node 1. In addition, some of the flavors of the final theory connected to node 2 are mesons that arise when node 3 is dualized. There is an ${\cal O}(N_2)$ excess in the numbers of branes at nodes 1 and 3. This is the fractional brane triggering the cascade and ultimately responsible for the end of the cascade via strong dynamics (the capping of the warped throat in the gravity dual).
%

The perturbative superpotential is just the one that follows from \eref{W_classic_F0_orientifolded} by reversing the direction of arrows and  the cyclic order in the trace and adding an overall minus sign
\beqa
W=-X^1_{32}X^1_{21}  X^2_{21}{}^TX^2_{32}{}^T+ X^2_{32}X^1_{21}X^1_{21}{}^T X^2_{32}{}^T + X^1_{32} X^2_{21}X^2_{21}{}^T X^1_{32}{}^T \, .
\label{W_classic_F0_before_bottom}
\eeqa
For simplicity, and since they do not participate in the dynamics we are interested in, here and in the remaining analysis of this model we will not consider the flavors. 

Node 1 has $N_f=N_c+4$ and hence confines, disappearing from the quiver, and generates a 1-instanton superpotential 
\beq
W_{\rm gauge\, inst} \sim {\rm Pf} \, M_{22} \, ,
\label{W_node_1_F0}
\eeq 
with $M_{22} $ the matrix of mesons for node 1 
\beq
M_{22}=\left(\begin{array}{cc} M^{11}_{22} \ & M^{12}_{22} \ \\[.1 cm] M^{21}_{22} & M^{22}_{22} \end{array} \right) =\left(\begin{array}{cc} X^1_{21} X^1_{21}{}^T \ & X^1_{21} X^2_{21}{}^T \ \\[.1 cm] X^2_{21} 
X^1_{21}{}^T & X^2_{21} X^2_{21}{}^T \end{array} \right) \, . 
\label{M22_F0}
\eeq 
Note that this matrix is antisymmetric due to insertions of the invariant tensor $J$ in the $USp$ color index contractions.

Notice that \eref{W_node_1_F0} is not the desired coupling \eref{pf-one}, which instead involves mesons of node 3. For clarity, we will reserve the calligraphic font for the operators generated by D-brane instantons. Combining \eref{W_node_1_F0} and \eref{W_classic_F0_before_bottom} expressed in terms of the mesons $M^{ij}_{22}$, we obtain the following superpotential
\beqa
W={\rm Pf} \, M_{22}  - M^{12}_{22} X^2_{32}{}^T X^1_{32}+M^{11}_{22} X^2_{32}{}^T X^2_{32}+M^{22}_{22}X^1_{32}{}^TX^1_{32} \, .
\eeqa
Next, dualizing node 3 we arrive at the theory we were originally interested in, whose quiver is shown in \fref{quiver_F0_instanton}. This dualization generates the following mesons of node 3
\be
N_{22}=\left(\begin{array}{cc} N^{11}_{22} \ & N^{12}_{22}\ \\[.1 cm] N^{12}_{22} & N^{22}_{22}\end{array} \right)=\left(\begin{array}{cc} X^1_{32}{}^T X^1_{32} \ & X^1_{32}{}^T X^2_{32} \ \\[.1 cm] X^2_{32}{}^T X^1_{32} & X^2_{32}{}^T X^2_{32} \end{array} \right) \, ,
\label{N22_F0}
\eeq
and inverts the direction of all arrows connected to node 3. This is not the full meson matrix since there are mesons involving the flavors. These mesons are responsible for flipping the direction of the arrows connecting the D7-branes to node 2, but do not show up in the non-perturbative term generated by the instanton, so we ignore them here. Note that $N_{22}$ from \eref{N22_F0} should not be confused with ${\cal N}_{22}$ from \eref{cal_N22_F0}. Both of them are mesons of node 3, but in different theories: $N_{22}$ is made out of fields in the quiver of \fref{quiver_F0_last_step} (after confining node 1), while ${\cal N}_{22}$ is made out of fields in \fref{quiver_F0_instanton}. 

In terms of these mesons, the superpotential becomes
\beqa
W& =& {\rm Pf} \, M_{22}  - M^{12}_{22} N^{21}_{22}+M^{11}_{22} N^{22}_{22} +M^{22}_{22}N^{11}_{22} \nonumber \\ 
& + &  N^{21}_{22} X^{1}_{23} X^{2}_{23}{}^T- N^{22}_{22} X^{1}_{23} X^{1}_{23}{}^T-N^{11}_{22} X^{2}_{23} X^{2}_{23}{}^T \, . 
\label{W_F0_3}
\eeqa
The $M^{ij}_{22}$ and $N^{ij}_{22}$ mesons are massive, and can be integrated out using their equations of motion, i.e. vanishing of their F-terms. From $\partial W/\partial N^{ij}_{22}=0$ we obtain
\beq
M^{11}_{22} =  X^{1}_{23} X^{1}_{23}{}^T  \ \ \ \ \ M^{22}_{22} =  X^{2}_{23} X^{2}_{23}{}^T  \ \ \ \ \ M^{12}_{22} =  X^{1}_{23} X^{2}_{23}{}^T\ .
\eeq
Recalling (\ref{cal_N22_F0}), we have that $M^{ij}_{22}=\mathcal{N}^{ij}_{22}$. Plugging this into \eref{W_F0_3}, we precisely obtain the D-brane instanton contribution \eref{pf-one}. 

It is straightforward to include the D7-brane flavors throughout the analysis and check that the flavor superpotential is correctly reproduced.
%

\subsection{A Useful Trick}

\label{section_useful_trick}

In the above analysis we did not have to deal with the complications of using the F-terms of fields entering the Pfaffian superpotential. However, such operations are necessary in more complicated examples.
In this section we introduce a simple trick to recast Pfaffian superpotentials, which allows to perform these operations in a straightforward manner, with results which amount to combinatorial operations in the periodic quiver/dimer. This trick provides an explanation of the formal process of dualizing empty nodes.

Let us consider a $USp(N_c)$ theory with $N_f$ chiral multiplets in the fundamental, and no superpotential, for $N_f=N_c+4$. The theory confines and develops a superpotential $W\sim {\rm Pf} M$ for its mesons $M=QQ$. The trick is to rewrite this contribution as a path integral over a set of auxiliary Grassmann variables $\lambda_i$, transforming under the flavor group, and with a cubic coupling to the mesons $\lambda M\lambda^T$. Many field theory computations regarding F-terms can be carried out at the level of this description, and the eventual integration over the Grassmann variables reconstructs the Pfaffian at the end. 

\begin{figure}[!ht]
\begin{center}
\includegraphics[width=11cm]{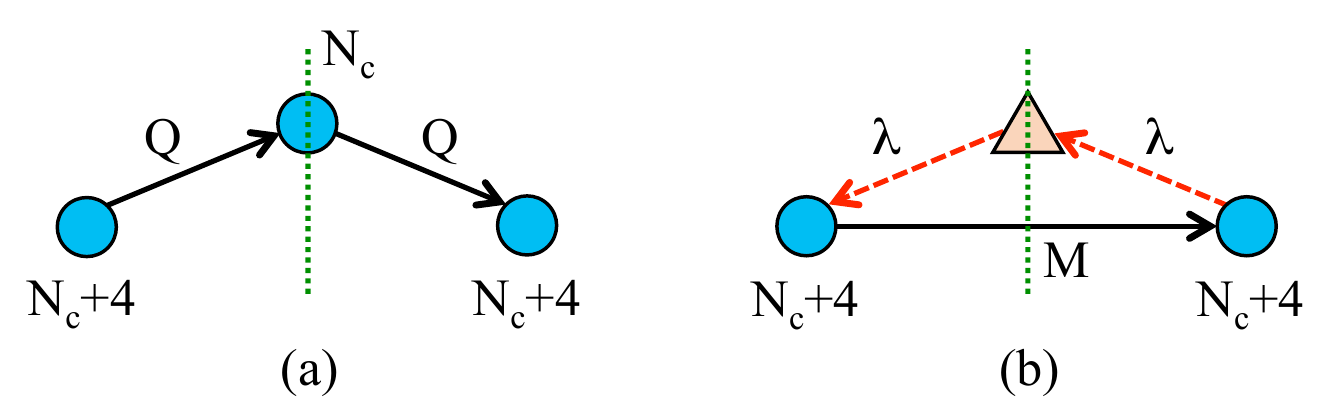}
\caption{(a) Quiver diagram for a $USp(N_c)$ electric theory with $N_f=N_c+4$. (b) The magnetic dual, with the empty node supporting an instanton, whose zero modes are shown as dashed arrows.} 
\label{trick}
\end{center}
\end{figure}

Before discussing explicit examples, we would like to mention that this description is closely related to the magnetic description of this theory: in quiver language, it contains an empty node which supports an $O(1)$ instanton, with fermion zero modes coupling as dictated by the magnetic quiver, and which agree with the properties of the above auxiliary Grassmann variables. A graphical representation of the trick, which thus amounts to a formal Seiberg duality, is shown in \fref{trick}.

 Let us apply the trick to the UV gauge theory of the earlier section, with the   quiver shown in  \fref{quiver_F0_last_step}. As above, node 1 confines and generates the Pfaffian superpotential \eref{W_node_1_F0} for its mesons. This can be rewritten by introducing a set of Grassmann variables $\lambda^1_{12}$ and $\lambda^2_{12}$, transforming in the antifundamental of $U(N_2)$. Their coupling to the mesons of (\ref{M22_F0}) can be formally combined with the original tree level superpotential, leading to
 \beqa
 &&  + \lambda^1_{12} M_{22}^{12} \lambda^2_{12}{}^T-\lambda^1_{12} M_{22}^{11} \lambda^1_{12}{}^T - \lambda^2_{12} M_{22}^{22}\lambda^2_{12}{}^T \nonumber \\
&&-X^1_{32} M_{22}^{12} X^2_{32}{}^T+X^1_{32} M_{22}^{11} X^1_{32}{}^T + X^2_{32} M_{22}^{22}X^2_{32}{}^T \, .
 \eeqa
This is just as dictated by the quiver of the Seiberg dual theory. Dualizing now node 3 and defining its mesons $N$ as in \eref{N22_F0}, we have
 \beqa
&&   + \lambda^1_{12} M_{22}^{12} \lambda^2_{12}{}^T-\lambda^1_{12} M_{22}^{11} \lambda^1_{12}{}^T - \lambda^2_{12} M_{22}^{22}\lambda^2_{12}{}^T  \\
&& -M_{22}^{12} N_{22}^{21}+ M_{22}^{11} N_{22}^{11} +  M_{22}^{22}N_{22}^{22}+ N_{22}^{21}X^1_{23}X^2_{23}{}^T- N_{22}^{11}X^1_{23}X^1_{23}{}^T- N_{22}^{22}X^2_{23}X^2_{23}{}^T \, . \nonumber
 \eeqa
Upon integrating out the massive fields, the F-terms of $N_{22}^{ij}$ impose $M_{22}^{ij}=X^i_{23}X^j_{23}{}^T$. The superpotential becomes precisely \eref{fzm-supo}. It is then manifest that integration over the auxiliary Grassmann variables finally reconstructs the superpotential of the D-brane instanton \eref{pf-one}.

The lesson is that confinement and the introduction of the Pfaffian superpotential for $USp$ nodes with $N_f=N_c+4$ is, via this trick, equivalent to performing the Seiberg duality leaving the node with the instanton and the corresponding charged zero modes. This effectively brings the duality cascade one step down towards the IR theory. This idea will be useful in the analysis of more involved examples in coming sections.

\bigskip

\section{Non-Cascading Geometries: $\mathbb{C}^3/\mathbb{Z}_3$ Examples}
\label{sec:non-cascading}

The second class of examples we want to consider involves theories that naively do not cascade. By this we mean theories that do not cascade, even if we consider the {\it full} quiver associated to the singularity and allow arbitrary ranks for all its nodes and the addition of flavors. Prototypical examples of this situation are provided by e.g.  $\mathbb{C}^3/(\mathbb{Z}_n \times \mathbb{Z}_m)$ chiral orbifolds, and orientifolds thereof. Earlier works studying other aspects of D-brane instantons on orbifolds of $\mathbb{C}^3$ can be found in \cite{Bianchi:2007wy,Ibanez:2007tu,Buican:2008qe,Bianchi:2009bg}.

This problem of UV completing such systems was considered (for oriented quivers) in \cite{Cascales:2005rj} (see also \cite{Franco:2008jc} for applications and \cite{Franco:2005fd} for earlier related work), where it was shown that such theories can emerge after partial confinement in a more complicated quiver, which can now correspond to the IR limit of a duality cascade. In the dual gravity language, the UV geometry reduces to the IR one via a complex deformation, which for toric singularities can be described very easily using dimer diagrams \cite{GarciaEtxebarria:2006aq}. Hence, although the final geometry naively appears non-cascading, it can emerge from a cascade as a result of the same type of dynamics that works in more conventional examples.

\bigskip

\subsection{D-brane Instanton Couplings}
\label{sec:z3-instanton-couplings}

For concreteness we focus on a prototypical example, namely gauge theories arising from orientifolds of D3-branes (and possibly D7-branes) at the $\mathbb{C}^3/\mathbb{Z}_3$ singularity. This simple geometry gives rise to many interesting chiral theories, which have appeared in widely varied applications (see e.g. \cite{Lykken:1998ec,Aldazabal:2000sa,Cascales:2005rj} for some of them).\footnote{Here we consider the $4d$ $\mathcal{N}=1$ preserving orbifold of $\mathbb{C}^3$, which has geometric action $(1,1,-2)$ on the three complex coordinates. A globally consistent chiral Type I theory this kind of orbifold was originally constructed in \cite{Angelantonj:1996uy}. There is an additional SUSY preserving $\mathbb{C}^3/\mathbb{Z}_3$ orbifold. It has geometric action $(1,-1,0)$, preserves $4d$ $\mathcal{N}=2$ SUSY and hence gives rise to a non-chiral gauge theory.} 

\begin{figure}[!ht]
\begin{center}
\includegraphics[width=11.5cm]{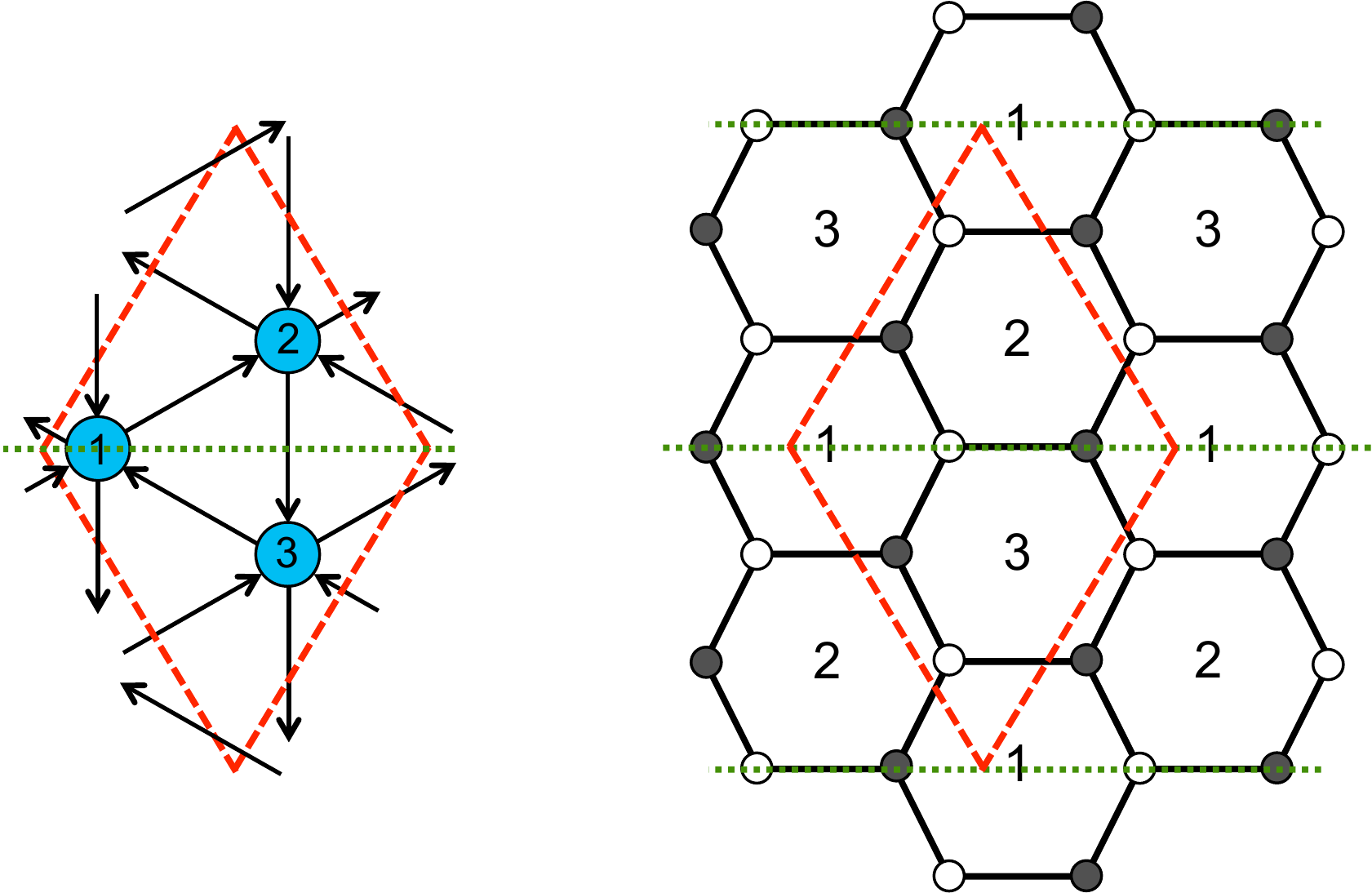}
\caption{The periodic quiver and dimer diagram for the orientifold of $\mathbb{C}^3/\mathbb{Z}_3$.}
\label{fig:quiver-dimer-sppz3}
\end{center}
\end{figure}

The periodic quiver and dimer of the $\mathbb{C}^3/\mathbb{Z}_3$ are shown in \fref{fig:quiver-dimer-sppz3}. There we have also shown an orientifold line that will be used later on. The superpotential of the theory without the orientifold is given by
\beqa
W=\epsilon_{ijk} X^i_{12}X^j_{23}X^k_{31} \, .
\eeqa
The superindices (which label the representation under an $SU(3)$ global symmetry) have a range $i=1,2,3$ and we use them to label the arrows going south, north-west, and north-east, respectively.

We consider the orientifold theory defined by the line reflection in the dimer. In order to obtain non-trivial superpotentials from D-brane instantons, we choose the orientifold charge such that the fixed gauge factor 1 is projected down to a $USp$ factor. Using the rules in \cite{Franco:2007ii}, this implies that the bifundamental  on top of the orientifold, which we denote by $X_{22}^1$, is projected down to a two-index antisymmetric of the $U(N_2)$. In addition, the two other bifundamentals $X^2_{22}$, $X^3_{22}$ are exchanged by the orientifold action, and produce one symmetric and one antisymmetric representations under $U(N_2)$. Other fields are related to their orientifold images by transposition, e.g. $X^2_{31}=X^3_{12}{}^T$, $X^1_{31}=X^1_{12}{}^T$, etc.

The superpotential of the orientifold theory reads (we express it in terms of fields above the orientifold line)
\beqa
W= X_{12}^3 X_{22}^1  X_{12}^3{}^T - X_{12}^2 X_{22}^1 X_{12}^2{}^T  -X_{12}^3 X_{22}^2 X_{12}^1{}^T + X_{12}^1 X_{22}^2 X_{12}^2{}^T  \, .
\label{W_perturbative_C3/Z3}
\eeqa
Here and in what follows $X_{22}^2$ is understood to split into a symmetric plus an antisymmetric representation.

\begin{figure}[!ht]
\begin{center}
\includegraphics[width=6cm]{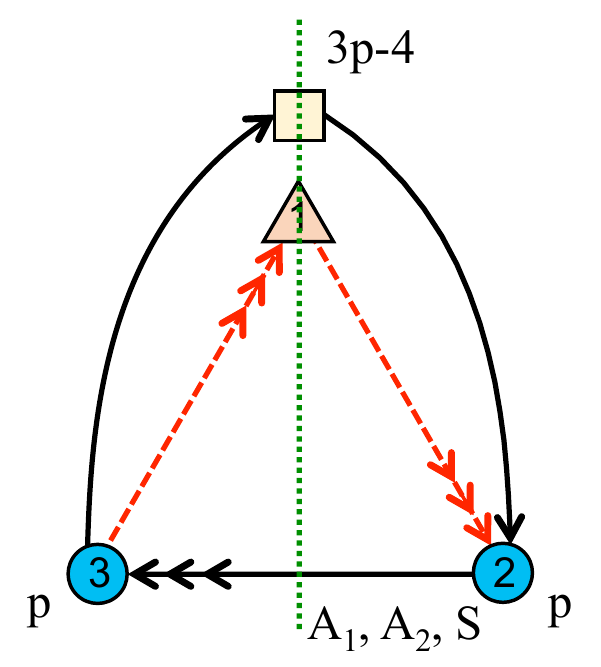}
\caption{The $\mathbb{C}^3/\mathbb{Z}_3$ orientifold quiver supporting a D-brane instanton. There are $p$ fractional branes, which cannot trigger a duality cascade, and a number of D7-branes to render the theory anomaly-free. Consistency requires $p$ to be even.}
\label{fig:c3z3-instanton-quiver}
\end{center}
\end{figure}

Let us focus on the theory with node 1 empty, shown in \fref{fig:c3z3-instanton-quiver}. This node can support an $O(1)$ instanton (thus with two neutral fermion zero modes) and a set of charged fermion zero modes depicted as dashed arrows in the figure. Even though the node is empty, we must cancel the Witten global anomaly of the $USp$ node, and demand that the number of incoming arrows is even, so $p$ is forced to be even in this model. In string theory, this requirement arises from cancellation of RR tadpoles in K-theory \cite{Uranga:2000xp}.

The fermion zero mode couplings can be obtained from the above superpotential by replacing chiral bifundamentals by zero modes, so we get
\beqa
 \lambda_{12}^3 X_{22}^1  \lambda_{12}^3{}^T - \lambda_{12}^2 X_{22}^1 \lambda_{12}^2{}^T  -\lambda_{12}^3 X_{22}^2 \lambda_{12}^1{}^T + \lambda_{12}^1 X_{22}^2 \lambda_{12}^2{}^T \, .
 \label{coupling-zero-modes}
\eeqa

Integrating the instanton partition function over these fermion zero modes we obtain the non-perturbative superpotential
\beqa
W_{\rm inst} \sim {\rm Pf}{\cal M} \, ,
\label{W_inst_C3/Z3}
\eeqa
where ${\cal M}$ is a matrix built out of the fields $X_{22}^i$. 

Since \eref{W_perturbative_C3/Z3} vanishes when node 1 is empty, the full superpotential is the sum of \eref{W_inst_C3/Z3} and the flavor superpotential. It is then given by
\beq
W=Q_{D7,2} A_1 Q_{D7,2}{}^T+{\rm Pf}\mathcal{M} \, ,
\label{full_W_C3/Z3}
\eeq
where the precise form of the flavor couplings is dictated by the embedding of the D7-branes, as explained in detail in \cite{Franco:2006es}, and we have used the notation in \fref{fig:c3z3-instanton-quiver} to emphasize we chose to couple the flavors to one of the antisymmetric fields. This theory is closely related to a model considered in \cite{Bianchi:2013gka}, as we explain in \sref{section_CFT_breaking}.

As already mentioned, this theory cannot be directly embedded in a UV duality cascade, since the $\mathbb{C}^3/\mathbb{Z}_3$ does not admit complex deformations (conversely, the gauge theory does not admit deformation fractional branes). In the following sections we study the embedding of this model into a different quiver theory (that reduces to this one by confinement), which in turn does admit a UV completion in terms of a duality cascade.

\subsection{The Parent Theory and its Complex Deformation}
\label{sec:parent-spp-z3}

It is convenient to first consider the unorientifolded parent theory to understand the basic ingredients to be used later for the orientifold model. The oriented $\mathbb{C}^3/\mathbb{Z}_3$ theory can be obtained from a richer quiver theory, a $\mathbb{Z}_3$ orbifold of the suspended pinch point  (or SPP/$\mathbb{Z}_3$ for short). Geometrically this occurs by a geometric transition inducing a complex deformation from SPP/$\mathbb{Z}_3$ to $\mathbb{C}^3/\mathbb{Z}_3$. From the field theory viewpoint, this process corresponds to partial confinement at the endpoint of a duality cascade, triggered by deformation fractional branes, which disappear in the IR $\mathbb{C}^3/\mathbb{Z}_3$ theory. 

\begin{figure}[!ht]
\begin{center}
\includegraphics[width=9cm]{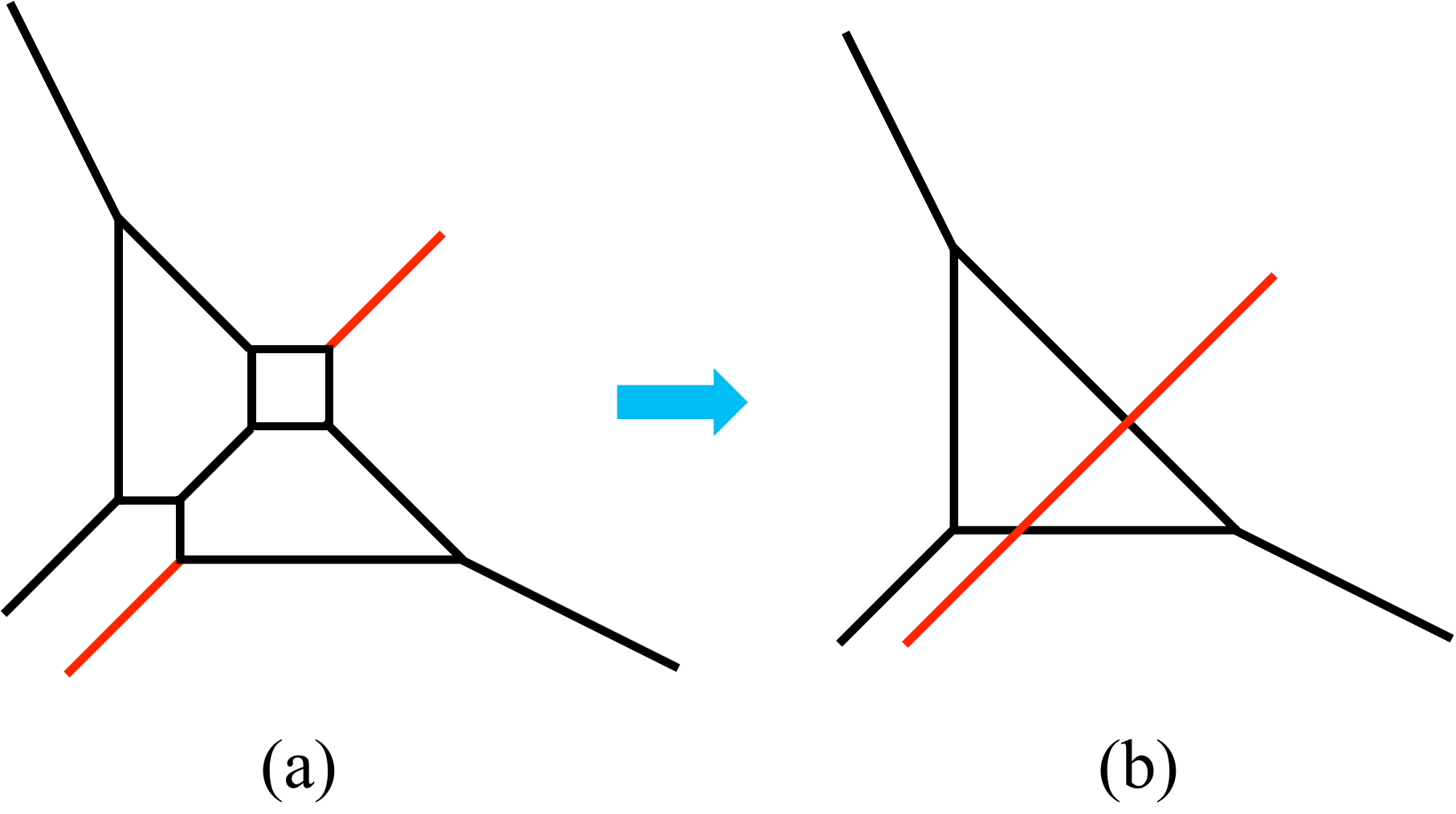}
\caption{Starting from the web diagram for SPP/$\mathbb{Z}_3$ (a) produces, via a complex deformation, the diagram for $\mathbb{C}^3/\mathbb{Z}_3$ (b).}
\label{fig:sppz3}
\end{center}
\end{figure}

Complex deformations of toric singularities are characterized very easily as splitting of the web diagram into subdiagrams \cite{Franco:2005fd}. This is illustrated for the deformation of SPP/$\mathbb{Z}_3$ to $\mathbb{C}^3/\mathbb{Z}_3$ in \fref{fig:sppz3}.

This construction for SPP/$\mathbb{Z}_3 \to \mathbb{C}^3/\mathbb{Z}_3$ has been extensively studied in \cite{Cascales:2005rj,Franco:2008jc} as a toy model in which a Standard Model-like theory emerges at the IR bottom of a throat. The possibility of a chiral theory at the end of the cascade follows from having a remnant singularity. In the following we review some aspects (and derive some new ones) that are most relevant for our present purposes.

\medskip

\begin{figure}[!ht]
\begin{center}
\includegraphics[width=6.5cm]{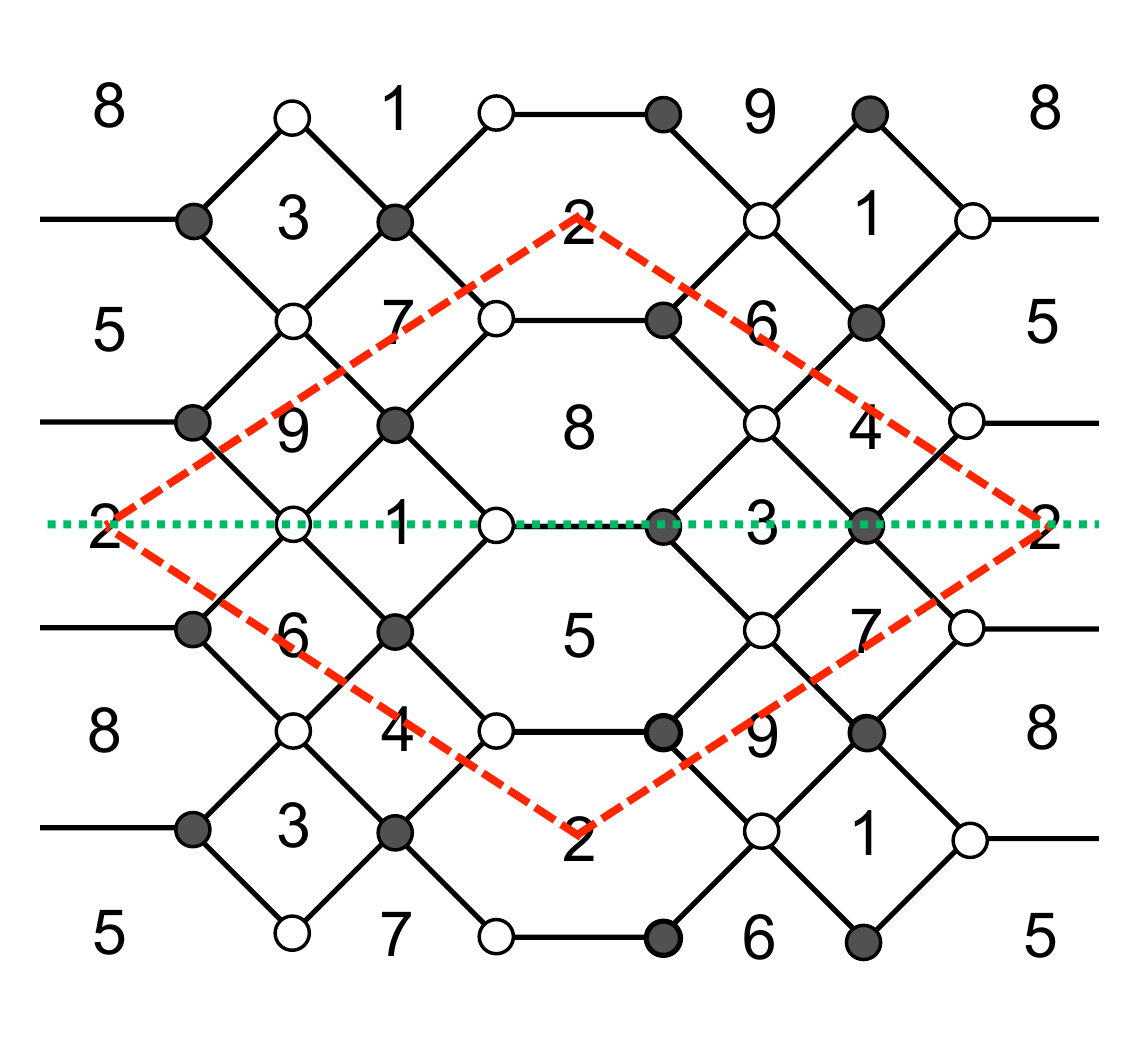}
\caption{The dimer for the SPP/$\mathbb{Z}_3$ theory. The green line is included for later use to describe the orientifold projection.}
\label{SPP_Z3_orientifold_fixed_line}
\end{center}
\end{figure}

The gauge theory on D3-branes over SPP was originally derived in \cite{Morrison:1998cs}. The gauge theory for D3-branes at SPP/$\mathbb{Z}_3$ can be obtained from it by standard orbifold techniques and is fully encoded in the dimer diagram shown in \fref{SPP_Z3_orientifold_fixed_line}. Every superpotential term of the theory is represented by a node in the dimer, or equivalently a plaquette in the dual periodic quiver, according to the dictionary introduced in \cite{Franco:2005rj}. Hence, we will follow the standard approach of dealing with dimer diagrams/periodic quivers for most of our discussion and only writing the explicit expressions for the superpotential that can be straightforwardly read from them when necessary. The choice of unit cell in  \fref{SPP_Z3_orientifold_fixed_line} is different from the one used in \cite{Cascales:2005rj, Franco:2008jc} for convenience to impose the orientifold action later on (already anticipated as the green line in the figure). Note that although the model is fully chiral, it is related to the non-chiral SPP   by orbifolding, as in \cite{Uranga:1998vf}. This is a recently rediscovered tool in the context of class $S_k$ theories \cite{Gaiotto:2015usa,Franco:2015jna,Hanany:2015pfa}.

Following \cite{Cascales:2005rj, Franco:2008jc}, the duality cascade of the SPP/$\mathbb{Z}_3$ theory and its deformation to the $\mathbb{C}^3/\mathbb{Z}_3$ theory corresponds, even in the presence of extra flavors,  to a $\mathbb{Z}_3$ orbifold of the cascade and complex deformation of the underlying SPP geometry to a smooth space, described in appendix \ref{sec:thespp}.  The cascade in SPP/$\mathbb{Z}_3$ is obtained by dualizing gauge factors in whole columns of squared faces in the dimer, which turn the squares in one of the adjacent columns into hexagons, and  the hexagons in the other adjacent column into squares. This has been recently studied in the context of class $S_k$ theories in \cite{Franco:2015jna,Hanany:2015pfa}. A similar pattern will hold in the orientifold theory, considered in the next section.

\subsection{The Orientifolded Theory and its Cascade}
\label{sec:unorient-spp-cascade}

Let us describe the orientifold of the SPP/$\mathbb{Z}_3$ theory we are interested in.\footnote{There are other possible orientifolds, with fixed points in the dimer diagram, but these are not compatible with the eventual complex deformation of SPP/$\mathbb{Z}_3$ to $\mathbb{C}^3/\mathbb{Z}_3$.} The orientifold is described as a line reflection in the dimer diagram, as shown in \fref{SPP_Z3_orientifold_fixed_line}. The orientifold action maps gauge groups 1, 2 and 3 onto themselves, while it identifies the pairs $4 \leftrightarrow 7$, $5\leftrightarrow 8$ and $6 \leftrightarrow 9$. We choose negative orientifold plane charge, in notation of \cite{Franco:2007ii}, so invariant gauge factors project onto $USp$ groups, and invariant fields onto two-index antisymmetric representations.

At this point it is possible to introduce flavors from D7-branes. Their behavior follows very closely the unorientifolded case, and for simplicity we omit them until the study of the IR behavior of the cascade in \sref{ssec:instanton}.

The resulting gauge theory thus has six gauge groups
\beq
USp(N_1) \times USp(N_2) \times USp(N_3)\times U(N_{4}) \times U(N_{5}) \times U(N_{6}) \ ,
\eeq
where to facilitate comparison, we have preserved the node labels in the parent theory. The rules to read out the spectrum and superpotential interactions are described in \cite{Franco:2007ii}. We simply mention that the bifundamental $(\antifund_5,\fund_8)$, which is mapped to itself under the orientifold, turns into a two-index (conjugate) antisymmetric of $U(N_5)$, $\antiasymm_5$. Fields not invariant under the orientifold, like e.g. $(\fund_3,\antifund_8)$, combined  with their image fields, e.g. $(\fund_5,\antifund_3)$, descend to bifundamentals of the corresponding group in the orientifold quotient. In these operations it is important to keep in mind that the orientifold maps gauge factors by relating their representations by conjugation, e.g. $U(N_4)\leftrightarrow U(N_7)$ such that the $\fund_4\leftrightarrow\antifund_7$.

\begin{figure}[!ht]
\begin{center}
\includegraphics[width=9cm]{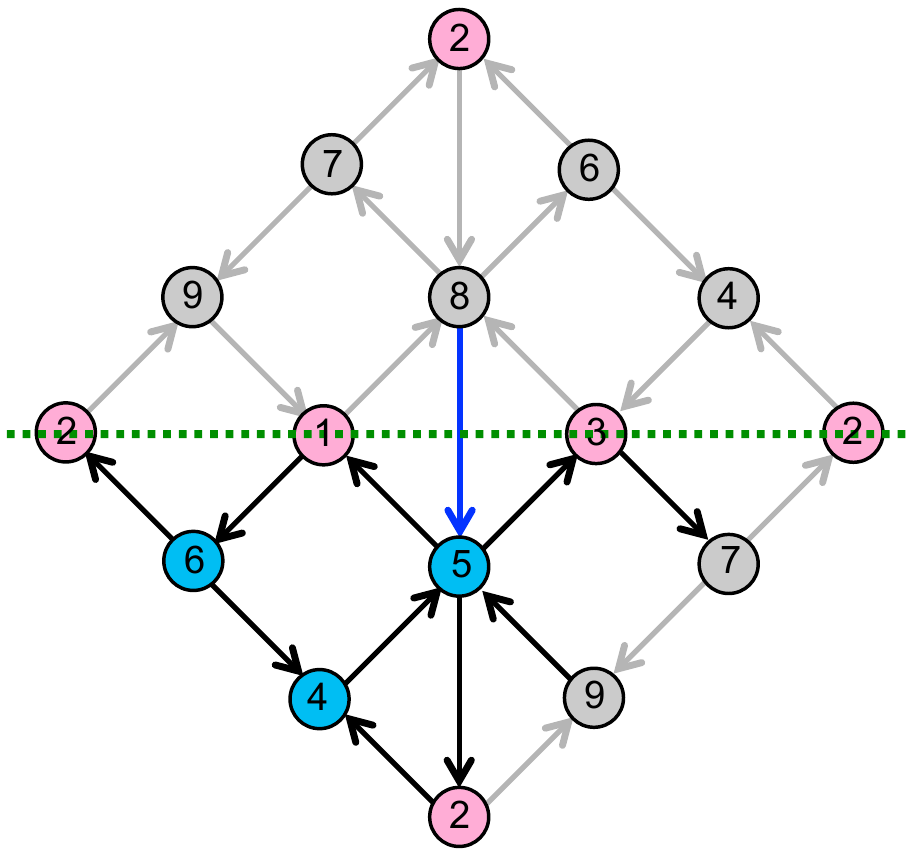}
\caption{The quiver diagram for the line orientifold of the  SPP/$\mathbb{Z}_3$ theory.}
\label{SPP_Z3_orientifold_fixed_line_quiver} 
\end{center}
\end{figure}

The final result is shown as a quiver in \fref{SPP_Z3_orientifold_fixed_line_quiver}. In order to simplify its interpretation, we color code nodes and arrows. Blue nodes are of $SU$ type (with their images in grey), while pink nodes are their own orientifold images and hence $USp$.  Black arrows from node $a$ to $b$ correspond to fields in the $(\fund_a,\antifund_b)$ (with their images depicted as light color arrows). The orientation of arrows at pink nodes is irrelevant, since $USp$ has not complex representations, but we preserve it as a useful bookkeeping device. Finally, the blue arrow transforms as $\antiasymm_5$. It is important not to double count these fields.

It is useful to recall that the SPP theory admits a T-dual \cite{Uranga:1998vf} in terms of a type IIA Hanany-Witten (HW) configuration \cite{Hanany:1996ie} of  two NS-branes (along 012345) and one rotated NS-brane (denoted as NS'-brane, along 012389), with 
D4-branes (alond 01236) suspended among them in the periodic coordinate $x^6$. In this picture, the orientifold we are describing corresponds to the introduction of an O8-plane (along 012345689) \cite{Feng:2001rh}. The orbifolded theory is realized by an additional $\mathbb{Z}_3$ orbifold of the HW configuration, acting in the directions 45, 89.

\bigskip

The discussion of the cascade is analogous to the parent SPP/$\mathbb{Z}_3$ theory without the orientifold. Namely, as each column of squares is preserved by the line orientifold, the cascade of Seiberg dualities is formally compatible with the orientifold. 
In a similar way to what we described in \sref{section_cascade_F0}, the original cascade sequentially dualizes the three nodes in each column, namely a period corresponds to the sequence (1,4,7) (2,5,8) (3,6,9). Due to the identification of nodes, in the orientifolded theory the cascade involves the two surviving nodes from each of these columns, namely: (1,4) (2,5) (3,6).

In the following we describe some details of the dualization procedure. A nice property of this particular example is that it avoids having to dualize nodes with matter in the antisymmetric representation.\footnote{The dual of  gauge theories with antisymmetric matter is only known in very few cases, see e.g. \cite{Berkooz:1995km,Pouliot:1995me,Brodie:1996xm, Terning:1997jj}.}

\subsubsection{Basic Step in the Duality Cascade}\label{ssec:SPP_cascade}

Let us discuss the dualization of nodes on a column of squares in the dimer in more detail. As in the unorientifolded case, we will see its effect is to shift the line of vertical arrows sideways. For concreteness, let us focus on the column consisting of nodes 1 and 4  (of type $USp$ and $SU$, respectively).

\bigskip

\subsubsection*{Dualizing Node 1}

\begin{figure}[!ht]
\begin{center}
\includegraphics[width=13.5cm]{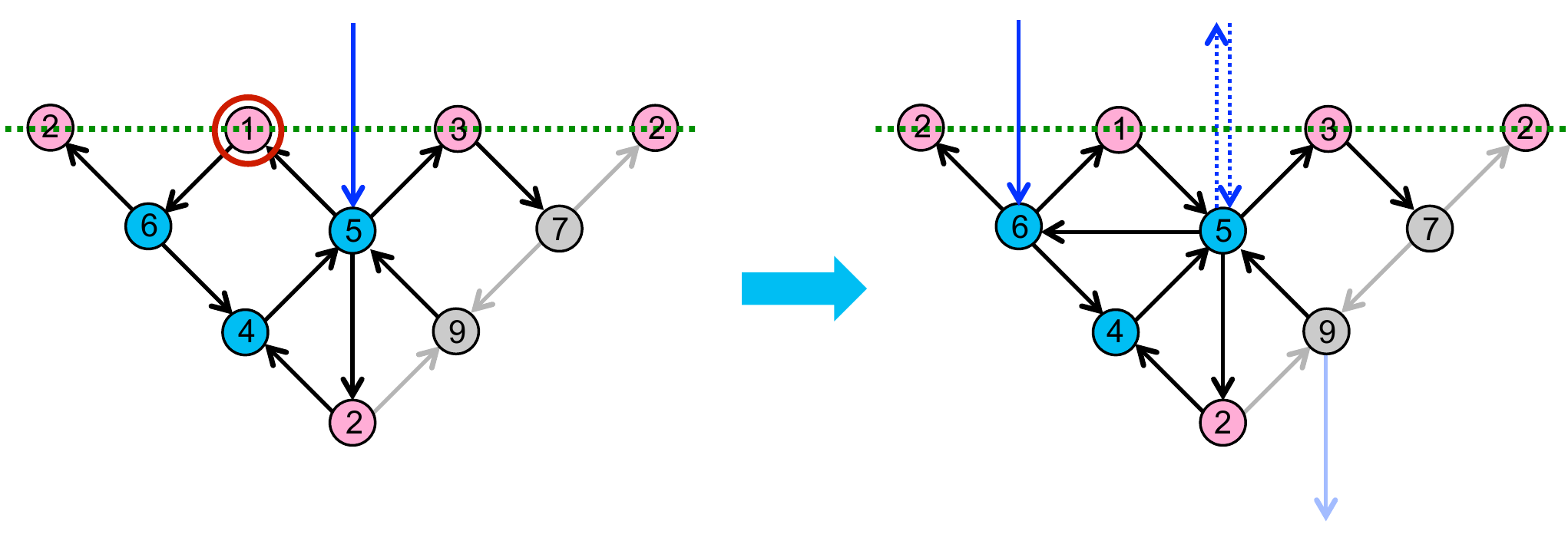}
\caption{Dualization of node 1, which is a $USp$ gauge group. In this figure and the ones that follow, it is important to keep in mind that in the orientifold, nodes 7 and 9 are identified with nodes 4 and 6, respectively.}
\label{SPP_Z3_orientifold_cascade_step_1}
\end{center}
\end{figure}

Consider dualizing the $USp$ type node 1, as shown in \fref{SPP_Z3_orientifold_cascade_step_1}.
The rules for Seiberg duality in quivers were described in \cite{Feng:2001bn}, and recast in terms of dimer diagram in \cite{Franco:2005rj}. Upon dualization, we invert the orientation of arrows connected to node 1, effectively introducing the dual quarks, which transform in conjugate representations under the symmetries of nodes 5 and 6. In addition, we introduce arrows corresponding to meson fields in representations $(\fund_5,\antifund_6)$, $\antiasymm_6$ and $\asymm_5$. The superpotential pairs up the latter with the original $\antiasymm_5$ in a mass term, and the pair can be integrated out. We have indicated these fields using dotted arrows to indicate that they disappear at low energies.

In \fref{SPP_Z3_orientifold_cascade_step_1}, the dark blue arrow on the right figure represents the  antisymmetric field charged under node 6, and the light blue one represents the same field at a different unit cell.

\bigskip

\subsubsection*{Dualizing Node 4}

\begin{figure}[!ht]
\begin{center}
\includegraphics[width=13.5cm]{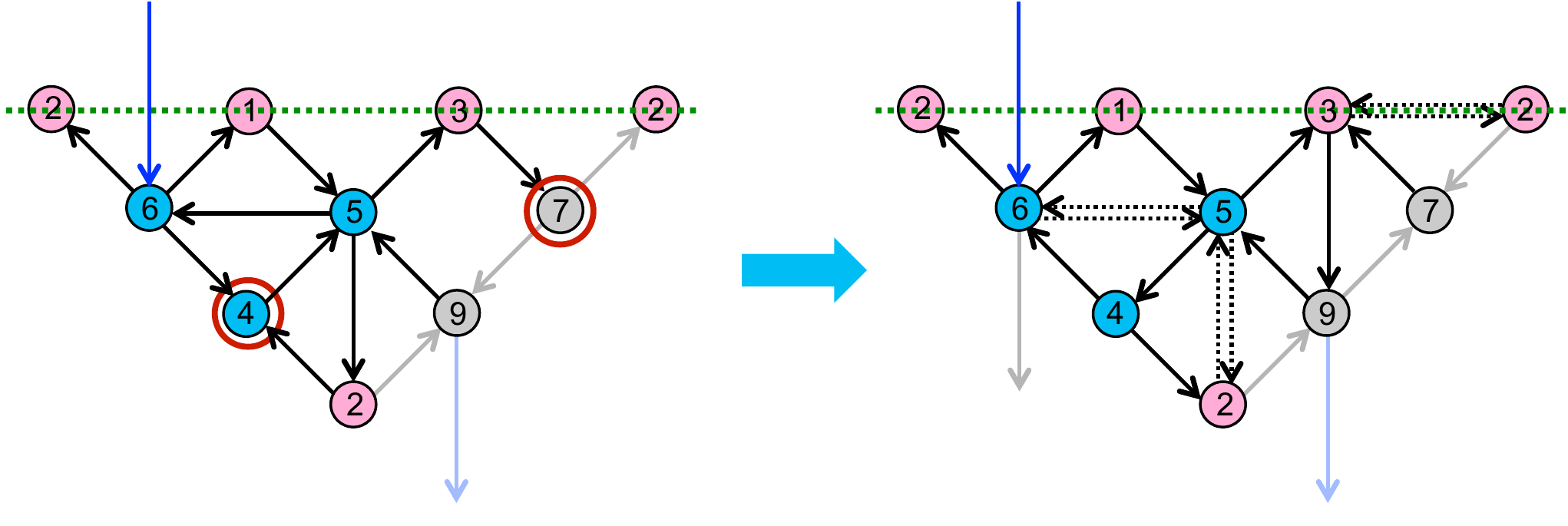}
\caption{Dualization of nodes 4 and 7, which are $SU$ gauge groups.}
\label{SPP_Z3_orientifold_cascade_step_2}
\end{center}
\end{figure}

In the second step, we dualize node 4 (and its orientifold image in the $\mathbb{Z}_2$ covering). This is an $SU$ gauge group and its dualization, which is rather standard, is shown in \fref{SPP_Z3_orientifold_cascade_step_2}. As before, dotted lines indicate pairs of field that become massive and can be integrated out. 

\begin{figure}[!ht]
\begin{center}
\includegraphics[width=6cm]{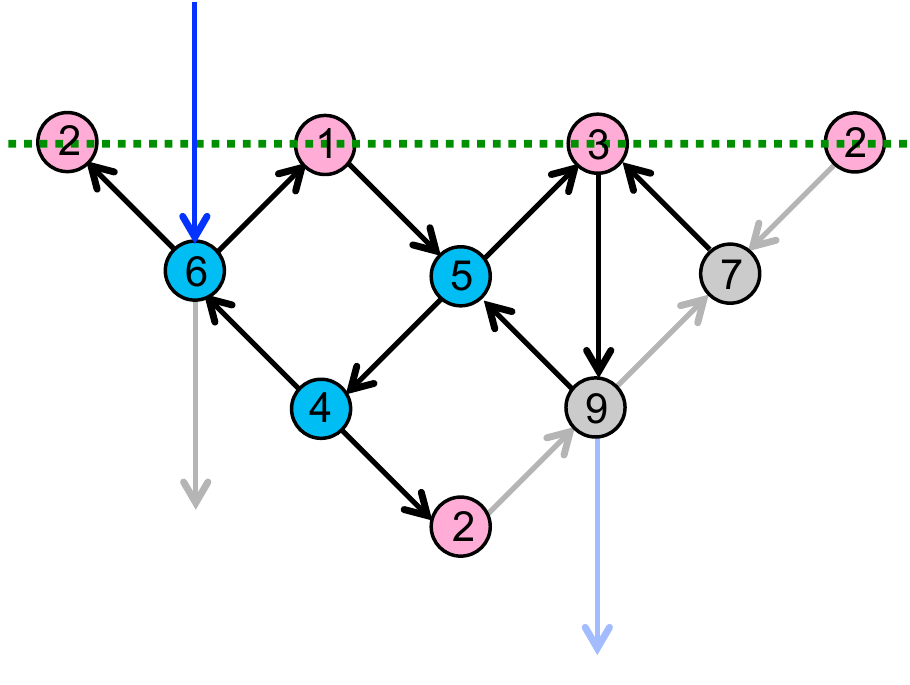}
\caption{Final result after dualizing the vertical column consisting of nodes 1 and 4.}
\label{SPP_Z3_orientifold_cascade_final}
\end{center}
\end{figure}

The final results is shown in \fref{SPP_Z3_orientifold_cascade_final}. The net effect of dualizing a column of nodes is a horizontal shift of the vertical arrows. An identical analysis applies to the dualizations of nodes (2,5) and (3,6), resulting in a periodic Seiberg duality cascade. To leading order in $1/N$, the change in ranks in as in the parent oriented theory, namely the reduction in the number of D3-branes is $N\to N-6M$. The precise numbers actually depend on the pattern of D7-brane flavors, and will be discussed in explicit examples below.

\bigskip

\subsection{The Instanton} 
\label{ssec:instanton}

We now choose a rank assignment and D7-brane content in the SPP/$\mathbb{Z}_3$ theory that relates, upon inclusion of strong dynamics, to the $\mathbb{C}^3/\mathbb{Z}_3$ orientifold theory introduced in section \sref{sec:z3-instanton-couplings}, and shown in \fref{fig:c3z3-instanton-quiver}. Basically, the strong dynamics of nodes 1, 4 and 7, which is responsible for the complex deformation, recombines the factors 2 and 3, and 8 and 9 (and their images 5 and 6), respectively. Hence, we set those numbers pairwise equal, and related to those of the resulting $\IC^3/\IZ_3$ theory, namely
\beqa
N_2=N_3=0 \quad , \quad N_5=N_6=N_8=N_9=p \, .
\eeqa
In order to verify the cancellation of anomalies, it is important to appropriately take into account the orientifold identifications of gauge groups and chiral fields. The confining nodes 1, 4 and 7 are taken to have a large number of branes, of order $M\gg p$. The precise values are obtained by demanding cancellation of anomalies, once we account for the D7-branes, which we locate on the edge 58.\footnote{For a discussion on how to describe D7-branes in terms of dimer models we refer the reader to \cite{Franco:2006es,Franco:2013ana}. D7-branes are described by paths traversing edges in the dimer model, which indicate the operator made out of D3-D3 chiral fields the flavors are coupled to in the superpotential.} We obtain
\beqa
N_1=M  \quad , \quad N_4=N_7=M+p \, .
\eeqa

\begin{figure}[!ht]
\begin{center}
\includegraphics[width=9cm]{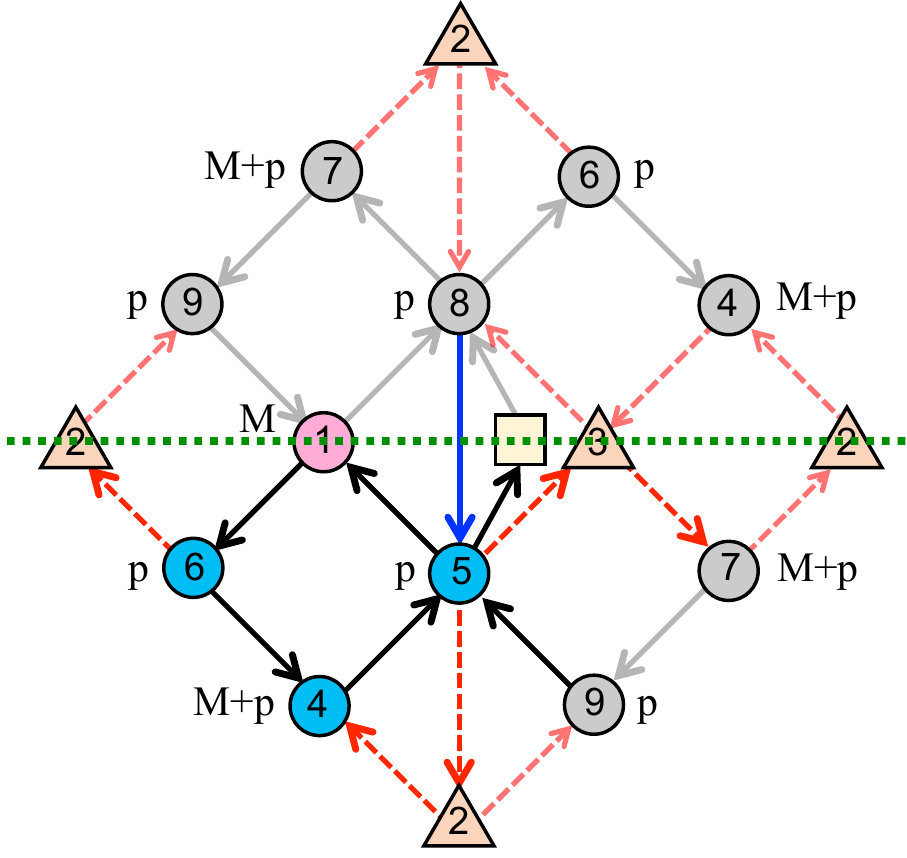}
\caption{Quiver for the orientifold of SPP$/\mathbb{Z}_3$ theory with an instanton. The configuration includes $3p-4$ flavor D7-branes, which are represented by the square node. Consistency requires $M$ and $p$ to be even.}
\label{sppz3-inst}
\end{center}
\end{figure}

The quiver right before the SPP$/\mathbb{Z}_3 \rightarrow \mathbb{C}3/\mathbb{Z}_3$ deformation is shown in Figure \ref{sppz3-inst}, including the $O(1)$ D-brane instantons at the empty nodes 2, 3. The requirement to cancel RR K-theory charge tadpoles \cite{Uranga:2000xp} (i.e. cancellation of Witten anomaly for the empty $USp$ nodes) requires $p$ even (as obtained already in the $\IC^3/\IZ_3$ orientifold) and $M$ even.

\begin{figure}[!ht]
\begin{center}
\includegraphics[width=15cm]{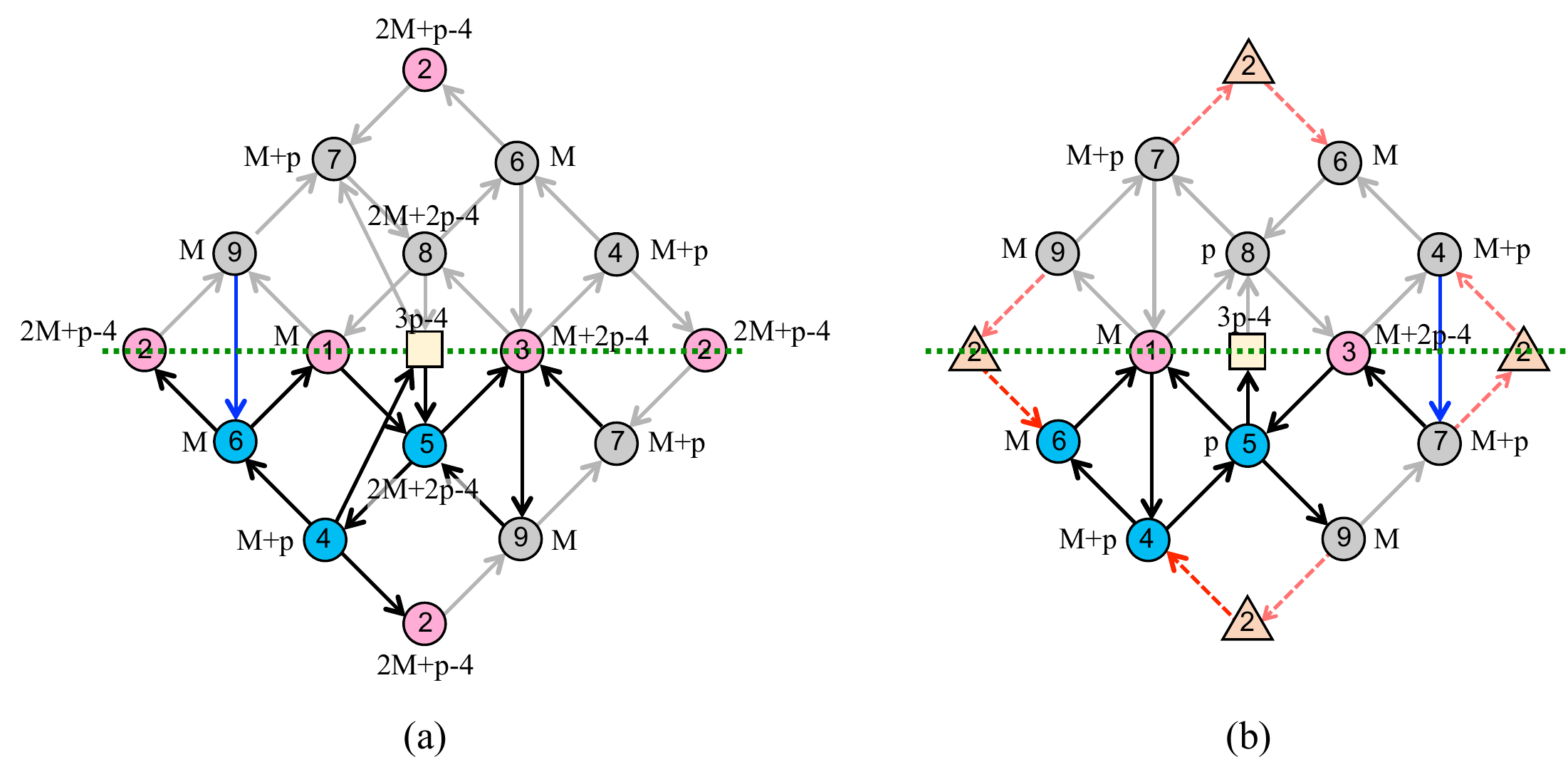}
\caption{(a) Quiver for the orientifold of SPP$/\mathbb{Z}_3$ theory two steps up the duality cascade. (b) Quiver obtained after considering the strong dynamics of node 2 and dualizing nodes 5 and 8. It corresponds to the SPP$/\mathbb{Z}_3$ orientifold in the next to last setp in the cascade.}
\label{fig:spp-uv-inst}
\end{center}
\end{figure}

We will later on derive the relation of this SPP/$\mathbb{Z}_3$ orientifold theory with the $\mathbb{C}^3/\mathbb{Z}_3$ orientifold theory in \fref{fig:c3z3-instanton-quiver}. Now we instead proceed to construct the UV gauge theory in which the D-branes instanton effects are realized as gauge instanton effects. This is done by moving up the cascade of SPP/$\mathbb{Z}_3$ until the empty nodes are filled. This occurs after two steps of dualization of whole columns, and the resulting quiver is shown in \fref{fig:spp-uv-inst}(a). The dualities are essentially as in the cascade in \sref{sec:unorient-spp-cascade}, with the only modification that we must include the D7-brane flavors in the discussion of some dualizations.

We now study the strong gauge dynamics of the theory in \fref{fig:spp-uv-inst}(a).
Consider node 2, which is a $USp$ group with $N_c=2M+p-4$, and its number of fundamentals is $N_f=2M+p$. Thus it has $N_f=N_c+4$ and generates a non-perturbative superpotential $W\sim {\rm Pf}\ M$ for its mesons $M$. Rather than adding this superpotential to the tree level one and using the F-terms by brute force, we prefer to take advantage of the trick in \sref{section_useful_trick}, and introduce a set of auxiliary Grassmann variables as arrows of opposite orientation to the original quarks, with cubic couplings to the mesons. As shown there, this is actually equivalent to performing a formal duality on node 2, leaving it as an empty node whose chiral multiplets are played by the Grassmann variables. In addition, we can dualize the nodes 5 and 8, to complete a duality in a whole column.  This operation can be carried out very easily in the quiver, as in the discussion in \sref{sec:unorient-spp-cascade}, and with the inclusion of D7-brane flavors in the dualization of nodes 5, 8. The resulting quiver is shown in \fref{fig:spp-uv-inst}(b).

\begin{figure}[!ht]
\begin{center}
\includegraphics[width=9cm]{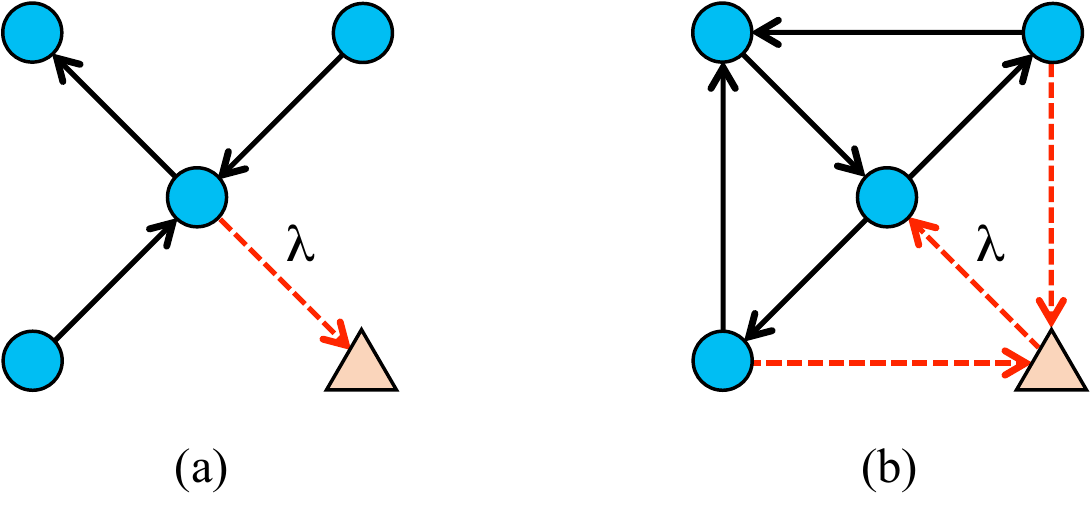}
\caption{Transformation of the fermion zero mode set upon Seiberg duality of a node under which they are charged.}
\label{fig:fzm-duality2}
\end{center}
\end{figure}

In this quiver, the node 3 is a $USp$ gauge factor with $N_c=M+2p-4$, and $N_f=M+2p$, hence it has $N_f=N_c+4$ so it confines and generates a non-perturbative superpotential $W\sim {\rm Pf} \ M'$ for its mesons. We can use again the trick in \sref{section_useful_trick} and introduce Grassmann variables as arrows of opposite orientation to the original quarks, with cubic coupling to the mesons. We must now dualize nodes 6 and 9, but this requires some care, since there are fermion zero modes of the instanton 2 charged under them. The change in the fermion zero mode spectrum is given in  \fref{fig:fzm-duality2}. It is easily obtained by demanding conservation of the net number of fermion modes on the corresponding instanton, and results in a formal duality reversing the Grassmann arrows and introducing `mesonic' ones.
 Applying this additional rule to the theory in \fref{fig:spp-uv-inst}(b), we get a quiver which corresponds precisely to \fref{sppz3-inst}. Clearly the trick to rewrite the Pfaffian superpotentials in terms of Grassmann variables makes it trivial to recover the right IR physics, as it effectively takes the theory one step down of the duality cascade.

\bigskip 

The next step is to consider the strong dynamics of the nodes 1, 4 and 7 in the quiver theory with two empty nodes, and show that after confinement it reduces to the $\mathbb{C}^3/\mathbb{Z}_3$ orientifold. Consider the theory in \fref{sppz3-inst} in the absence of the instantons on nodes 2, 3. The superpotential in this case is 
\beqa
W  =  Q_{5,{\rm D7}}{}^T X'_{55} Q_{5,{\rm D7}}+ X'_{55} X_{51}X_{51}^T  - X_{51}X_{16}X_{64}  X_{45}  \, ,
\eeqa
where the prime on $X'_{55}$ denotes that it transforms in the conjugate antisymmetric of node 5. The first term in the superpotential corresponds to the coupling of the flavors. The two other terms are obtained by restricting the SPP/$\mathbb{Z}_3$ superpotential to the non-empty nodes. It can be directly read from the surviving plaquettes in the periodic quiver shown in \fref{sppz3-inst}. In our regime of interest, $M\gg p$, the groups 1 and 4 (and its image 7) generate non-perturbative Affleck-Dine-Seiberg (ADS) superpotentials. We introduce the mesons
\beqa
 N=\begin{pmatrix} N_{55} & N_{56} \cr N_{65} & N'_{66}\end{pmatrix}=\begin{pmatrix} X_{51}X_{51}{}^T & X_{51}X_{16} \cr X_{16}{}^T X_{51}{}^T & X_{16}{}^T X_{16}\end{pmatrix}\quad ,\quad 
 {\tilde N}=X_{64}X_{45}\equiv {\tilde N}_{65} \ .
\eeqa
The superpotential for the theory in terms of these mesons is
\beqa
W & = & Q_{5,{\rm D7 }}{}^T X'_{55} Q_{5,{\rm D7 }}+ X'_{55} N_{55}-N_{56}{\tilde N}_{65} \nonumber \\ & + & M(\det {\tilde N})^{-\frac 1M}+(M-p+2)({\rm Pf}\ N)^{-\frac{2}{M-p+2}} \, ,
\eeqa
where we ignore the constants associated to the strong dynamics scale and some numerical prefactors.

The F-term for $N_{55}$ fixes $X'_{55}$ in terms of $\det N$, the F-terms for $N_{56}$ fixes ${\tilde N}_{65}$ in terms of $\det N$ and the F-term for ${\tilde N}_{65}$ fixes $N_{56}$ in terms of $\det {\tilde N}$. These mesonic vevs break the gauge symmetries of 5 and 6 (and those of the images 8 and 9) to their diagonal combination. Finally, the F-term for the antisymmetric  $X'_{55}$ sets $N_{55}= - Q_{5,{\rm D7}}Q_{5,{\rm D7}}{}^T$. The left over massless fields are the field $X_{65}$ (which decomposes as a ${\antisymm}_5+{\antiasymm}_5$) and $N'_{66}$ (which transforms as a ${ \antiasymm}_6$). 

The final superpotential can be read out by restricting to the simplest case, where the meson matrix $N$ of the $USp$ group is $4\times 4$, with entries $N_{ij}$ given by $2\times 2 $ blocks. The Pfaffian in this case reads
\beqa
\text{Pf } N \sim   N_{55}N'_{66} -  N_{56} N_{65}  \, ,
\eeqa
which allows us to write the F-term of $N_{55}$ as
\beqa
X'_{55} \sim ({\rm Pf}\ N)^{-\frac{2}{M-p+2}-1} N'_{66}  \, .
\eeqa
Using this we are left with a superpotential 
\beqa
W = Q_{5,{\rm D7}}{}^T N'_{66} Q_{5,{\rm D7}} \, , \label{supo-final-simple}
\eeqa
reproducing the perturbative piece of superpotential \eref{full_W_C3/Z3} for the $\mathbb{C}^3/\mathbb{Z}_3$ orientifold with flavors in \fref{fig:c3z3-instanton-quiver}, after the obvious map of fields.

\begin{figure}[!ht]
\begin{center}
\includegraphics[width=6.5cm]{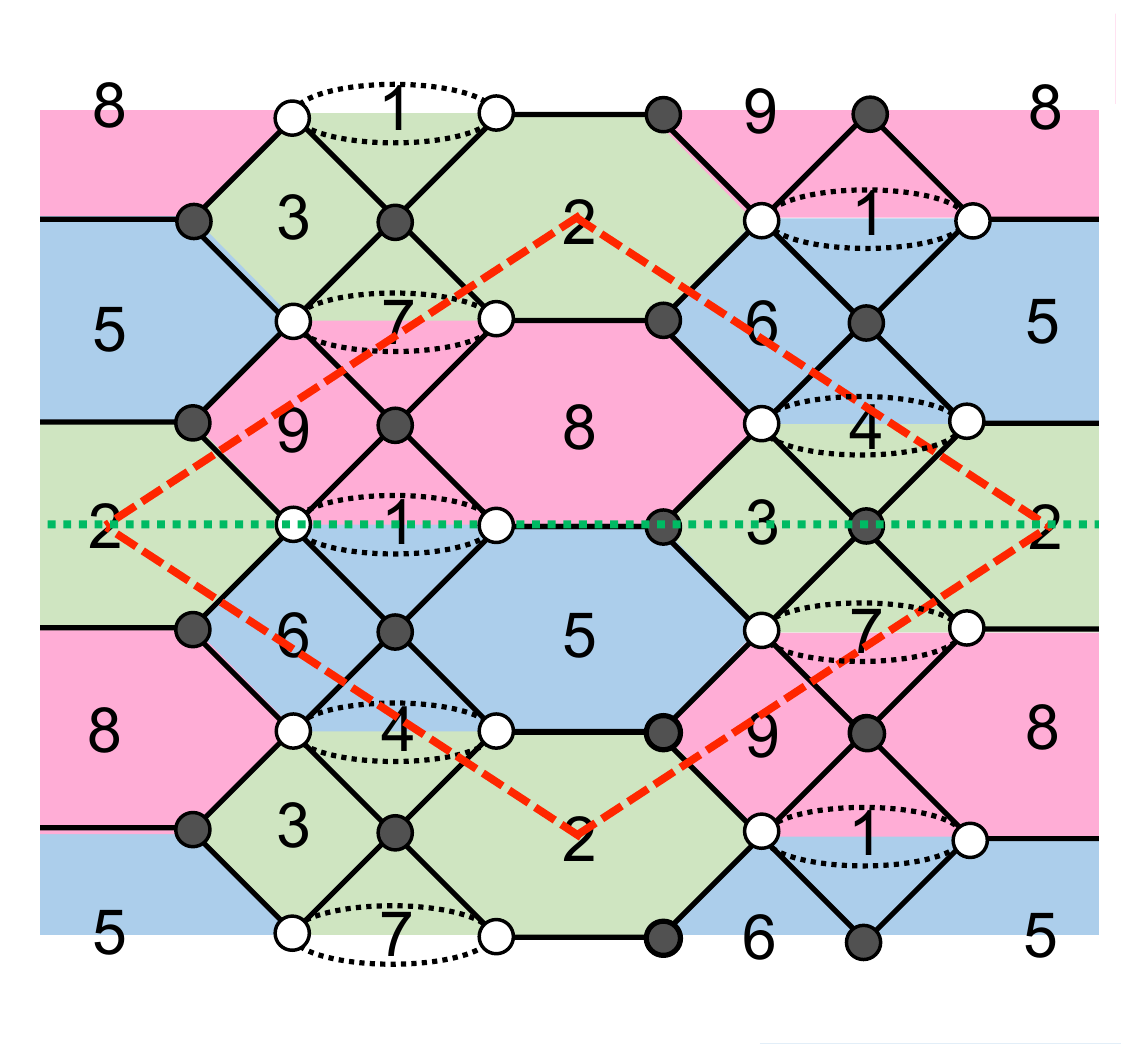}
\caption{The complex deformation of the geometry translates into strong gauge dynamics in the field theory. Nodes 1 and 4 (and its image 7) confine, whereas the pairs (2,3) and (5,6) (as well as its image pair (8,9)) are higgsed by non-zero expectation values of mesonic operators. The disappearance of part of the gauge symmetry due to confinement of nodes 1 and 4 is represented graphically by the shrinking and consequent disappearance of the corresponding faces in the dimer. This process is represented by the contraction of the dotted ovals. Higgsing translates into the combination of faces in the dimer.  We use pink, blue and green shades to identify the faces of the resulting dimer. As expected, they describe a line orientifold of $\mathbb{C}^3/\mathbb{Z}_3$.}
\label{deformsppz3}
\end{center}
\end{figure}

The strong dynamics responsible for the complex deformation we have just discussed can be nicely reproduced graphically in terms of the dimer as shown in \fref{deformsppz3}, using a procedure developed in \cite{Franco:2005zu,GarciaEtxebarria:2006aq}. The confinement of faces 1 and 4 (and its image 7) is represented by shrinking the corresponding faces, first into the dotted ovals and finally into a single edge. The disappearance of the faces is the dimer counterpart of the elimination of the associated gauge symmetry. In addition, as a result of these shrinking, the pairs (2,3) and (5,6) (as well as its image pair (8,9)) get recombined. This is precisely the pattern of higgsing triggered by the non-zero vevs we explained above. In \fref{deformsppz3} we indicate the faces that result from this process in color. The final unit cell has three hexagonal faces and an orientifold line, which identifies the blue and pink faces. This is precisely the dimer representation of the orientifold of $\mathbb{C}^3/\mathbb{Z}_3$ presented in \fref{fig:quiver-dimer-sppz3}. 

It is now easy to follow the fate of the instanton fermion zero modes of the complete quiver in \fref{sppz3-inst} in this confinement process. The fact that the node 4 (and its image 7) confines implies that fermion zero modes charged under them have to be saturated using gauge invariant couplings. In particular, the fermion zero modes $\lambda_{34}$ and $\lambda_{24}$ (and their images) must be saturated using the quartic coupling $\lambda_{34}\lambda_{24}{}^T \lambda_{24}\lambda_{34}{}^T$. This pattern of saturation of fermion zero modes reflects the fact that the two instantons act simultaneously in a 2-instanton process, very much in the spirit of \cite{GarciaEtxebarria:2007zv,GarciaEtxebarria:2008pi}. This is in agreement with the intuition that confinement of nodes 1, 4, 7 triggers the recombination of nodes 2 and 3. Once gauge groups 1, 4 and 7 confine, and after saturating the zero modes in the quartic coupling, the remaining fermion zero modes   are $\lambda_{53}$, $\lambda_{62}$ and $\lambda_{52}$, with couplings \footnote{Until here we took the convention of representing the sets of Grassmann variables as row vectors. From here on  we will change the convention since it is more convenient to write them as column vectors.}
\beqa
\lambda_{53}{}^T N'_{66}\lambda_{53} - \lambda_{62}{}^T N'_{66}\lambda_{62} - \lambda_{62}{}^T X_{65}\lambda_{52}  +\lambda_{52}{}^T X_{65} \lambda_{53} \, ,
\label{coupling-zero-modes2}
\eeqa
where as usual $X_{65}$ is understood to split into symmetric and antisymmetric parts.

This is precisely in agreement with the fermion zero modes of the instanton on the orientifold of $\mathbb{C}^3/\mathbb{Z}_3$, c.f. \fref{fig:c3z3-instanton-quiver}. Indeed, by comparing (\ref{coupling-zero-modes2}) with (\ref{coupling-zero-modes}), one sees that the structure of the couplings in both cases is the same, and it is easy to map the fields and zero modes of the SPP$/\mathbb{Z}_3$ theory to the $\mathbb{C}_3/\mathbb{Z}_3$ one. 

The conclusion is that the original non-perturbative gauge dynamics in the SPP$/\mathbb{Z}_3$ orientifold theory of two steps up the cascade will reproduce exactly the same non-perturbative superpotential as the  stringy D-brane instanton of the infrared $\mathbb{C}^3/\mathbb{Z}_3$ orientifold theory, given by \eref{W_inst_C3/Z3}. Combining this with \eref{supo-final-simple}, the full superpotential \eref{full_W_C3/Z3} is recovered.

\bigskip

\subsection*{Additional Remarks on the Complex Deformation}

The precise choice of ranks and flavors in the previous discussion was motivated by our goal of reproducing the theory in \fref{fig:c3z3-instanton-quiver} in the IR and understanding the emergence of D-brane instantons from gauge theory dynamics. As we explained, in this case the complex deformation is translated into higgsing triggered by ADS superpotentials.

It is interesting to mention that the complex deformation can be alternatively understood in terms of different strong dynamics. In particular, generalizing \cite{Klebanov:2000hb} (see e.g. \cite{Franco:2005fd}), it is possible to consider a setup with different ranks and flavors such that the fractional brane nodes do not generate ADS superpotentials but rather have complex deformed moduli spaces. On the mesonic branch, we precisely recover the theory of D3-branes probing the deformed geometry. Since our focus is on D-brane instantons, we will not discuss this interesting possibility any further.

\bigskip

\section{Models with Orientifold Points}
\label{sec:opoints}

The examples in the previous sections are based on orientifold quotients that have fixed lines in their action on the dimer diagram. In \cite{Franco:2007ii} there is a second kind of orientifold quotients, whose action on the dimer diagram has fixed points. It is easy to find chiral examples of this kind with duality cascades and deformations, providing a UV completion where D-brane instantons are realized as standard gauge instantons up in the cascade.
Since the analysis of these questions is very similar to our earlier explicit examples, we restrict the discussion to the construction of the basic dimer models.

It is easy to see that the SPP/$\mathbb{Z}_3$ singularity admits orientifolds with fixed points in the dimer, but they are not compatible with the complex deformation to $\mathbb{C}^3/\mathbb{Z}_3$. At the level of the gauge theory, the deformation fractional branes used in the previous section are not mapped to themselves. Instead, they map to a second kind of deformation fractional brane. Introducing both kinds of fractional branes however does not lead to a complex deformation; rather, it is equivalent (by the addition/removal of regular D3-branes) to introducing $\NN=2$ fractional branes, whose strong dynamics does not give complex deformations, but enhancon phenomena \cite{Johnson:1999qt}.

In the Hanany-Witten (HW) T-dual, orientifolds with fixed points in the dimer correspond to O6-planes (e.g. in the directions 0123789), sitting at opposite points in the $x^6$ direction. For the SPP, the only orientifold invariant configuration of the two NS- and one NS'-brane is to have the NS'-brane stuck on top of an O6, and have the two NS-branes away forming a $\mathbb{Z}_2$ invariant pair. The two kinds of deformation fractional branes are D4-branes stretched between the NS'- and one of the two NS-branes. There is no $\mathbb{Z}_2$ invariant way to recombine one NS- and one NS'-brane to account for the T-dual of the complex deformation. 

\medskip

\begin{figure}[!ht]
\begin{center}
\includegraphics[width=6cm]{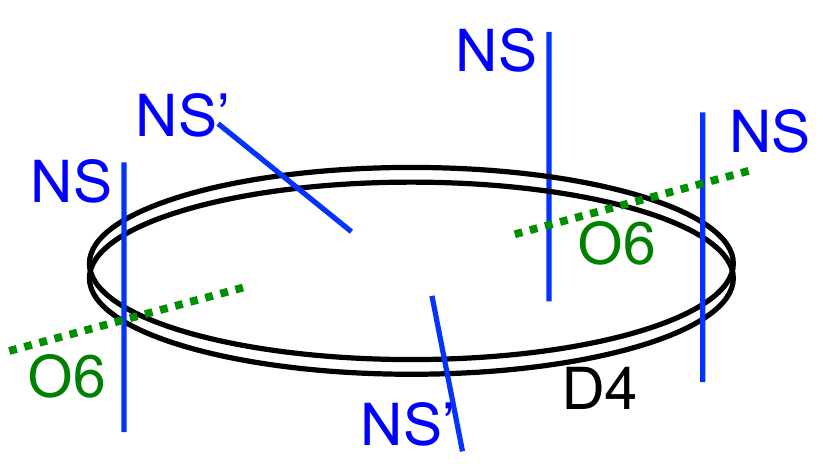}
\caption{Hanany-Witten configuration of NS5-branes, D4-branes and an O6-plane that is T-dual to D3-branes at an orientifold of the real cone over $L^{2,3,2}$.}
\label{HW_orientifold_L232}
\end{center}
\end{figure}

Using this HW perspective, it is however easy to device a different model admitting an orientifold quotient with fixed points in the dimer, and compatible with the complex deformation. Consider a HW T-dual with three NS-branes and two NS'-branes, and locate one NS-brane on top of an O6-plane, and two (NS,NS') pairs away from it, forming a $\IZ_2$ invariant system. This configuration is shown in \fref{HW_orientifold_L232}. This configuration admits a recombination of the NS- and NS'-branes in a pair (accompanied by the recombination of the orientifold image pair). In the picture of D3-branes at toric singularities, the geometry is $xy=z^3w^2$, see \cite{Uranga:1998vf}, and it admits a complex deformation to $\mathbb{C}^3$ compatible with the orientifold. Now, we can perform a $\IZ_3$ quotient of these systems, as in the recently introduced class $S_k$ theories \cite{Franco:2015jna}, and obtain a theory that admits a complex deformation to $\mathbb{C}^3/\mathbb{Z}_3$, compatible with the orientifold.

\begin{figure}[!ht]
\begin{center}
\includegraphics[width=7.5cm]{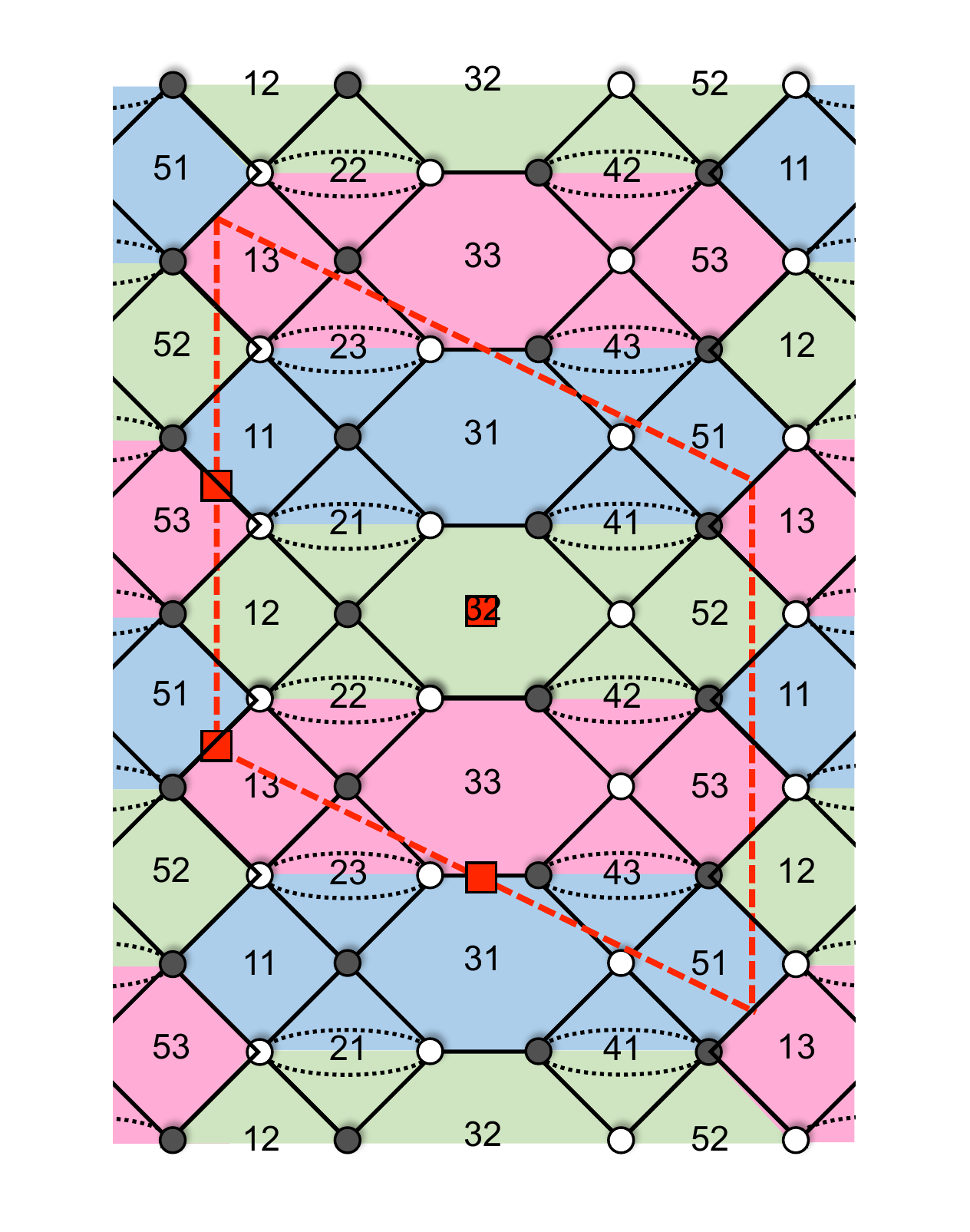}
\vspace{-.2cm}\caption{The dimer for a fixed point orientifold of the $\mathbb{Z}_3$ quotient of $xy=z^3w^2$. The faces 21, 22, 23, 41, 42 and 43 correspond to the deformation fractional brane, and form an invariant set under the orientifold action. The deformation is depicted as shrinking those faces as indicated by the dotted ovals. The pink, blue and green shades correspond to the faces of the resulting dimer after the complex deformation, and describe a point orientifold of $\mathbb{C}^3/\mathbb{Z}_3$.}
\label{fig:dimer-opoint}
\end{center}
\end{figure}

In the picture of D3-branes at toric singularities, the corresponding geometry is a $\IZ_3$ quotient of $xy=z^3w^2$, which is also known as the real cone over $L^{2,3,2}$ \cite{Franco:2005sm}.
It is straightforward to construct the dimer diagram describing the gauge theory, which is shown in \fref{fig:dimer-opoint}. The picture also displays the fixed points of the orientifold action, which is clearly compatible with the complex deformation to the $\mathbb{C}^3/\mathbb{Z}_3$ theory, because the fractional branes (corresponding to faces  21, 22, 23, 41, 42, 43) form an invariant set under the orientifold action. The final dimer faces after complex deformation are depicted as three hexagons with different background colors).

There is one face of the original dimer that is mapped to itself under the orientifold action. If empty, and for suitable choice of orientifold charges, it can support instantons that contribute to the superpotential. Up in the cascade, when the corresponding face is filled, the effect should be generated by a standard gauge instanton. In the infrared, some nodes confine and force non-trivial mesonic vevs which recombine certain faces and reconstruct the dimer for $\mathbb{C}^3/\mathbb{Z}_3$. By an analysis similar to that in earlier sections, it is easy to show that the gauge non-perturbative effects of the UV theory flow to the D-brane instanton effects of the infrared $\mathbb{C}^3/\mathbb{Z}_3$ theory. We refrain from a more detailed discussion of these and other similar examples, which can easily constructed with the techniques we have presented.

\bigskip

\section{Flavoring the Non-Perturbative Superpotential}
\label{section_CFT_breaking}

In this section we consider an alternative flavor configuration for the class of models studied in \sref{sec:non-cascading}. An interesting application of D-brane instantons has been proposed in \cite{Bianchi:2013gka}, where it was argued that certain 4d $\NN=1$ superconformal gauge theories arising from D3-branes at orientifold singularities actually suffer a non-perturbative breaking of superconformal invariance by D-brane instanton effects. This occurs because the gauge theories are realized in terms of quivers that contain empty nodes with formally `non-zero beta-functions'; these do not produce gauge factors, but can support D-brane instantons whose contribution to the effective action is weighted by a non-trivial scale. The example we consider is one of the simplest theories that have been argued to realize this scenario. Independently of this application, this theory is also interesting because it exhibits a novel feature: a D-brane instantons generates a non-perturbative superpotential involving the flavors.

In this section, we provide a UV embedding of one such example in \cite{Bianchi:2013gka}, based on an orientifold of $\mathbb{C}^3/\mathbb{Z}_3$, using a duality cascade of an orientifold of the SPP/$\mathbb{Z}_3$ theory, in which the IR D-brane instanton is realized as a standard gauge instanton at some higher step up in the cascade.

\begin{figure}[!ht]
\begin{center}
\includegraphics[width=6cm]{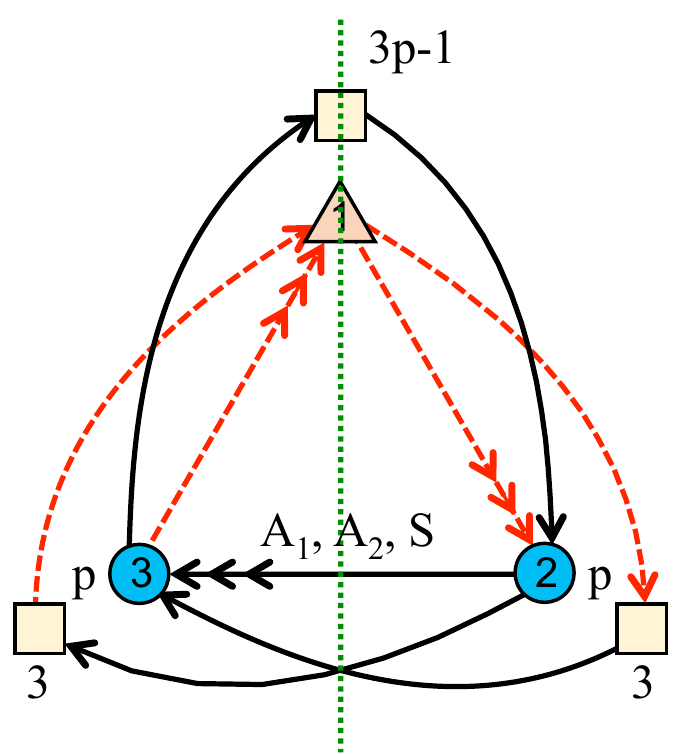}
\caption{The quiver for the orientifold of $\mathbb{C}^3/\mathbb{Z}_3$ theory with flavors. Consistency requires $p$ odd in this case.}
\label{cft-z3}
\end{center}
\end{figure}
  
%
The example we consider is the orientifold of $\mathbb{C}^3/\mathbb{Z}_3$ is the last model  in Table 3 of \cite{Bianchi:2013gka}, whose quiver we reproduce in \fref{cft-z3} in our notation for convenience. The gauge group is $SU(p)$, the D7-brane $SU(3)\times SO(3p-1)$ behaves as a global symmetry, and the fields transform as
\beqa
&SU(p)_{\rm D3}\times SU(3)_{\rm D7}\times SO(3p-1)_{\rm D7}&  \nonumber\\
&(\symm;1,1)+2(\asymm;1,1)+ (\fund; \fund,1) + (\antifund;1,\fund)& .
\eeqa
This model can be realized using the orientifold of $\mathbb{C}^3/\mathbb{Z}_3$ in \sref{sec:non-cascading}. Notice that in this case the RR K-theory charge cancellation conditions require odd $p$ instead.
The dimer diagram for this theory is the one in  \fref{fig:quiver-dimer-sppz3}, and the SPP$/\mathbb{Z}_3$ orientifold theory providing its UV completion is shown in \fref{SPP_Z3_orientifold_fixed_line}.  

The setup in \fref{cft-z3} is very similar to the one presented in \sref{sec:z3-instanton-couplings}, \fref{fig:c3z3-instanton-quiver}, with the 
only difference being the inclusion of extra stacks of three flavor branes in the regular representation of the orbifold (thereby neither contributing to the net anomaly nor to the RR tadpoles). Using this relation, it is easy to find a UV completion of this theory by following the analysis in \sref{sec:non-cascading}. We keep the discussion brief and omit some of the details that were already given in the previous analysis.

Note from \fref{cft-z3} that the 3 additional D7-branes can couple to any of the three types of charged zero modes discussed in the end of \sref{ssec:instanton}, that we dubbed $\lambda_{53}$, $\lambda_{62}$ and $\lambda_{52}$. In what follows we focus on one of these cases, namely coupling the flavor branes to $\lambda_{52}$. The other cases can easily be analyzed by following the same ideas explained in this section, and give rise to different non-perturbative terms that we discuss by the end of this section.

\begin{figure}[!h]
\begin{center}
\includegraphics[width=9cm]{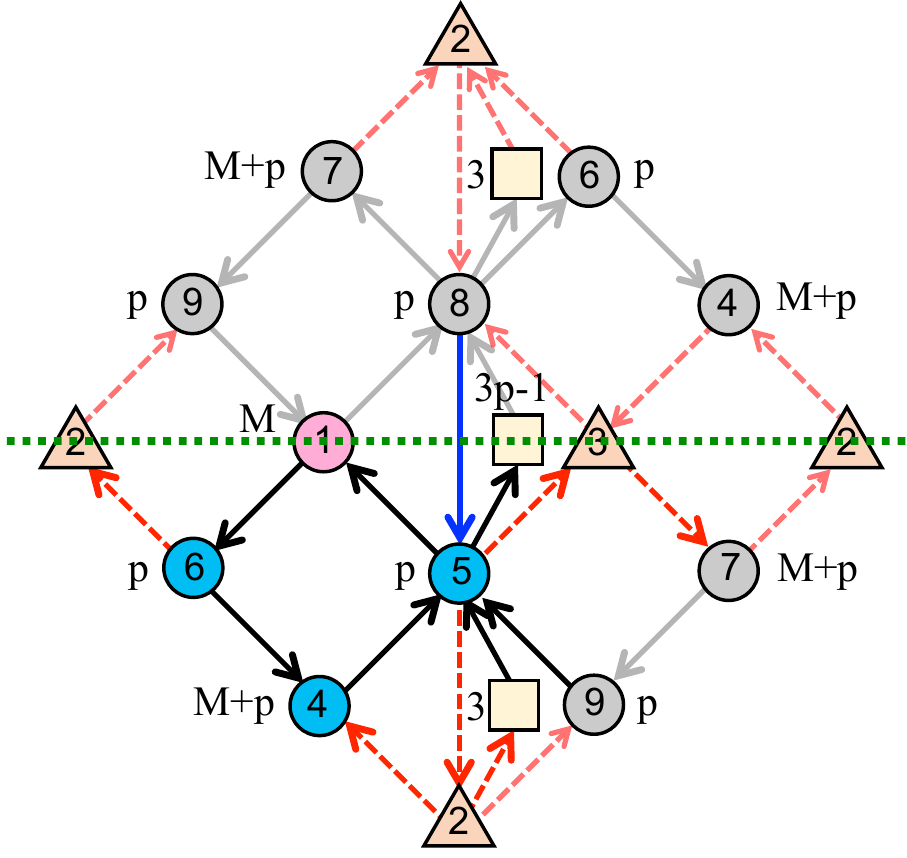}
\hspace*{-10pt}
\caption{Periodic quiver for the orientifold of the SPP$/\mathbb{Z}_3$ theory with extra flavor branes and D-brane instantons sitting on nodes 2 and 3. Consistency requires $p$ odd and $M$ even.}
\label{fig:spp-ir-inst-bianchi}
\end{center}
\end{figure}

Given that the IR theory only differs from \fref{fig:c3z3-instanton-quiver} by the introduction of three regular D7-branes, it is natural to construct the UV theory as the  SPP$/\mathbb{Z}_3$ orientifold therein, with three additional regular D7-branes.  The corresponding quiver for the case we are considering is shown in \fref{fig:spp-ir-inst-bianchi}. There is a stack of $(3p-1)$ D7-branes on top of the orientifold plane (as compared with $(3p-4)$ in \fref{sppz3-inst}), and three D7-branes coupled to nodes 5 and 2  (and their image D7-branes). Note that the D3-brane numbers remain as in \fref{fig:c3z3-instanton-quiver}. Cancellation of K-theory RR charge requires $p$ odd (as already obtained for the $\mathbb{C}^3/\mathbb{Z}_3$ orientifold) and $M$ even.

\begin{figure}[h!]
\begin{center}
\includegraphics[width=15cm]{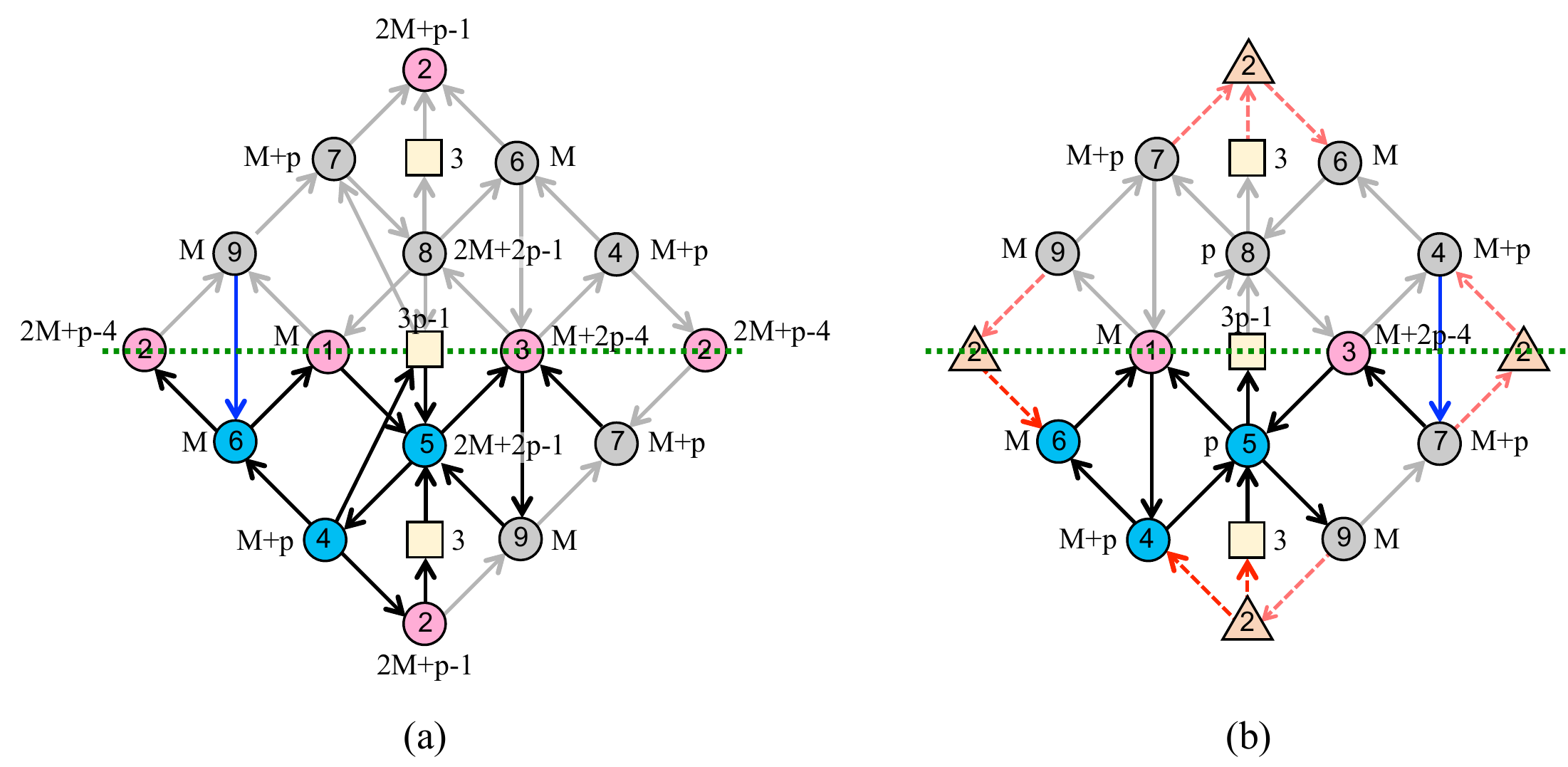}
\hspace*{-30pt}
\caption{(a) Periodic quiver for the orientifold of SPP$/\mathbb{Z}_3$ theory with the extra flavor brane stacks that leads to the setup in \fref{cft-z3} in the IR.  (b) The same periodic quiver obtained after considering the strong dynamics of node 2 and dualizing node 5 and its image 8. The triangle on node 2 represents the stringy instanton.}
\label{fig:spp-uv-inst-flav}
\end{center}
\end{figure}

In order to find a UV theory in which the D-brane instanton effects arise from standard gauge dynamics, we must move up the duality cascade until the nodes 2 and 3 are filled. This is done by applying Seiberg duality to whole columns, as described in \sref{sec:unorient-spp-cascade}. Since the new regular D7-branes do not introduce flavors for the nodes 3, 6 and 9, the result of the first duality up the cascade is given by \fref{fig:spp-uv-inst-flav}(b), with the addition of 3 regular D7-branes. The next step up the cascade requires dualizing the nodes 2, 5 and 8, which receive new flavors from the D7-branes. The resulting theory is given by \fref{fig:spp-uv-inst-flav}(a), with the 3 additional regular D7-branes, and with the numbers of D3-branes on the dual nodes increased by 3 units for nodes 5, 8 and  2 when compared to \fref{fig:spp-uv-inst}.

It is clearly straightforward to continue dualizing up to the UV to reconstruct a duality cascade where the inclusion of the flavor branes only produce $\mathcal{O}(p/N)$ corrections, with $N$ the number of regular D3-branes.

We can use by now familiar arguments to show that the strong gauge dynamics of the theory in \fref{fig:spp-uv-inst-flav}(a) reproduces the D-brane instanton effect in the final $\mathbb{C}^3/\mathbb{Z}_3$ orientifold.
First, node 2 is a $USp$ gauge group with $N_f=N_c+4$ that confines and generates a Pfaffian non-perturbative term in the superpotential. By using the trick in \sref{section_useful_trick}, the inclusion of this term is equivalent to performing a Seiberg duality on node 2. Dualizing also nodes 5 and 8, we get the quiver diagram of \fref{fig:spp-uv-inst-flav}(b).  
 
At this point, node 3 is a $USp$ group with $N_f=N_c+4$, so it confines. Using again the trick in \sref{section_useful_trick}, we end up dualizing node 3. Then we dualize nodes 6 and 9 taking into account the transformation of fermion zero modes in \fref{fig:fzm-duality2}. The resulting theory is precisely \fref{fig:spp-ir-inst-bianchi}.

Finally, the last step in the RG flow is the strong dynamics of nodes 1 and 4 (and its image 7). This process triggers the complex deformation that takes the theory from the orientifold of SPP$/\mathbb{Z}_3$ to the orientifold of $\mathbb{C}^3/\mathbb{Z}_3$. Since the D7-branes do not introduce flavors for these nodes, this process is as described at the end of \sref{ssec:instanton}. At this point it is important to note that the only difference between the quivers in \fref{fig:spp-ir-inst-bianchi} and \fref{sppz3-inst} is that the former contains extra flavor branes that give rise to the new coupling
\beqa
\lambda_{2,{\rm D7'}}Q_{{\rm D7'},5}\lambda_{52} \, ,  
\label{extra-coupling}
\eeqa
where the prime in the D7 flavor index represents that this stack of D7 branes is different from the one we had in  the setup of \sref{sec:non-cascading}, which gave rise to the superpotential term (\ref{supo-final-simple}).
This coupling was not present in the theory we analyzed in \sref{sec:non-cascading}. Fortunately, since this coupling contains no field or zero mode charged under groups 1 and 4 (and its image 7), it is a mere spectator of the deformation process. Therefore, we can just borrow the analysis in \sref{sec:non-cascading} and add the extra coupling in the end. This leaves a final superpotential given by (\ref{supo-final-simple}) and the couplings
\beqa
\lambda_{53}{}^T N'_{66}\lambda_{53} - \lambda_{62}{}^T N'_{66}\lambda_{62} - \lambda_{62}{}^T X_{65}\lambda_{52}  +\lambda_{52}{}^T X_{65} \lambda_{53}  + \lambda_{52}\lambda_{2,D7'} Q_{D7',5} \ .
\label{couplings1}
\eeqa
This is precisely one of the possible sets of couplings described by the quiver in \fref{cft-z3}. The other possibilities arising from that quiver differ from this one in coupling the new stack of three D7-branes to $\lambda_{53}$ or $\lambda_{62}$ instead of $\lambda_{52}$. In these cases the effect of this new stack on the IR is generating a new coupling of the form of (\ref{extra-coupling}), involving the desired charged zero mode. This leaves a series of couplings for e.g. $\lambda_{53}$:
\beqa
\lambda_{53}{}^T N'_{66}\lambda_{53} - \lambda_{62}{}^T N'_{66}\lambda_{62} - \lambda_{62}{}^T X_{65}\lambda_{52}  +\lambda_{52}{}^T X_{65} \lambda_{53}   + \lambda_{53}\lambda_{3,D7'} Q_{D7',5} \ . 
\label{couplings2}
\eeqa
This series of couplings leads to a different non-perturbative term compared to (\ref{couplings1}) once we saturate the zero mode integral over the instanton partition function. Noting that the non-perturbative term will be of the form (\ref{W_inst_2}), this means that the matrices on the Pfaffian arising from (\ref{couplings1}) and (\ref{couplings2}) will be different and will have different Pfaffians. For completeness, we mention that coupling the three D7-branes to $\lambda_{62}$ leads to the same non-perturbative term as the one obtained from (\ref{couplings2}). These two cases generate different matrices, but their Pfaffians are the same. This is a consequence of the dimer in \fref{fig:quiver-dimer-sppz3} being left-right symmetric, which implies that coupling the D7-branes to a zero mode on the left (i.e. $\lambda_{62}$) has the same effect as coupling them to the zero mode on the right (i.e. $\lambda_{53}$).
 
\bigskip

\section{Conclusions}

\label{section_conclusions}

We have presented the first UV completions in terms of duality cascades and gauge dynamics of D-brane instanton effects in chiral gauge theories arising on D-branes at singularities. Previously existing examples only concerned non-chiral theories. Our examples also include the first gauge theory completions of D-brane instantons couplings involving flavors. Although we focused on instantons contributing to the superpotential, it is clear that the results are general, and apply also to other instantons correcting other terms of the 4d effective action.

We studied both cascading and non-cascading IR geometries. The latter do not admit fractional branes and hence, naively, do not allow duality cascades. This obstacle can be bypassed by obtaining them by complex deformations of UV geometries that support cascades. We have discussed in detail how the complex deformations translate into strong dynamics of the field theories. An interesting aspect is that, since the UV and IR theories are related by recombination of gauge factors, the IR instanton is reproduced by a multi-instanton process in the UV theory.

In general, deriving D-brane instanton couplings from gauge theory dynamics involves the intricacies of integrating out fields participating in Pfaffian terms in the superpotential using their F-terms. In \sref{section_useful_trick} we introduced a simple trick that recasts Pfaffian contributions as path integrals over auxiliary Grassmann variables. With such rewriting, Seiberg dualities and all other necessary manipulations of the gauge theories reduce to the standard ones for general ranks of the gauge groups, i.e. for ranks that do not give rise to non-perturbative superpotentials. These transformations can be efficiently implemented in terms of the dimer or periodic quiver. The auxiliary variables precisely correspond to the fermionic zero modes in the D-brane construction.

We have primarily presented our ideas in terms of explicit examples. However, we do not foresee any obstruction preventing them from applying to general orientifolds of toric CY$_3$ singularities.

\bigskip

\section*{Acknowledgments}

We would like to thank D. Galloni and A. Mariotti for collaboration at early stages of this project. We are also indebted to M. Bianchi, M. Montero, J. F. Morales and G. Zoccarato for useful discussions. S. F. would like to thank the Aspen Center for Physics (NSF grant 1066293) for hospitality during the final stages of this project. A. R. and A. U. are partially supported by the grants FPA2012-32828 from the MINECO, the ERC Advanced Grant SPLE under contract ERC-2012-ADG-20120216-320421 and the grant SEV-2012-0249 of the ``Centro de Excelencia Severo Ochoa" Programme. 

\bigskip

\appendix

\section{The SPP Cascade and Orientifolds Thereof}
\label{sec:thespp}

Our examples in the main text are based on (orientifolds of) the SPP$/{\mathbb Z}_3$ singularity. Several of their properties can be discussed in the parent SPP (or orientifolds thereof), before orbifolding. In this appendix we review some of these parent systems.

\subsection{The SPP, its Cascade and its Complex Deformation}
\label{sec:spp-oriented}

\begin{figure}[!ht]
\begin{center}
\includegraphics[width=8cm]{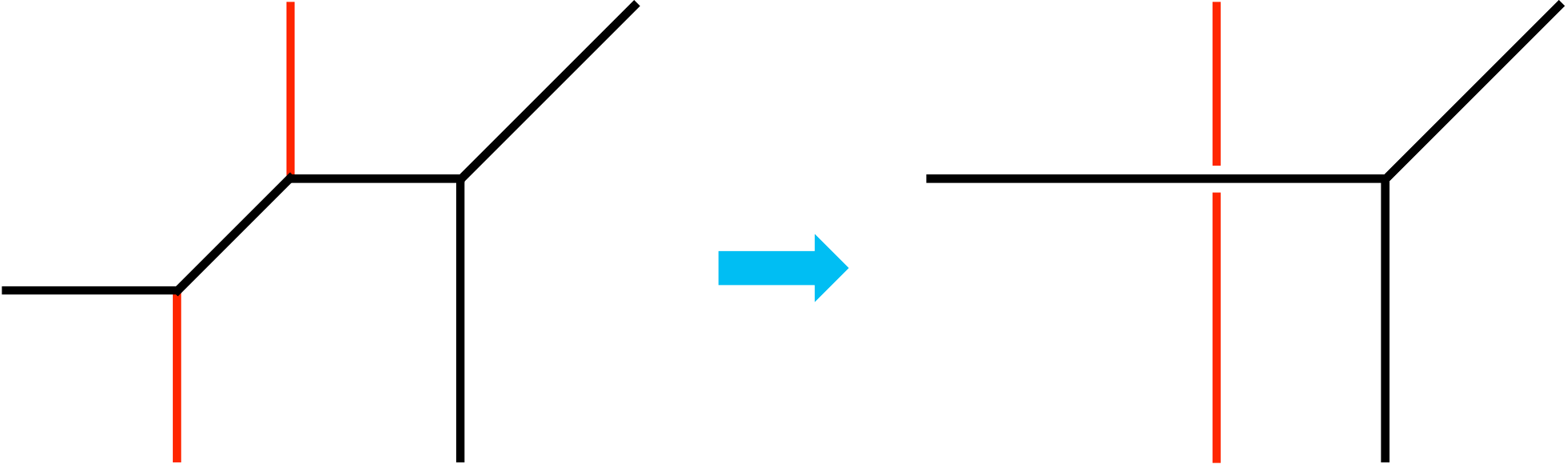}
\caption{Web diagram for the SPP and its deformation to a smooth geometry.}
\label{web_SPP}
\end{center}
\end{figure}

\begin{figure}[!ht]
\begin{center}
\includegraphics[width=12cm]{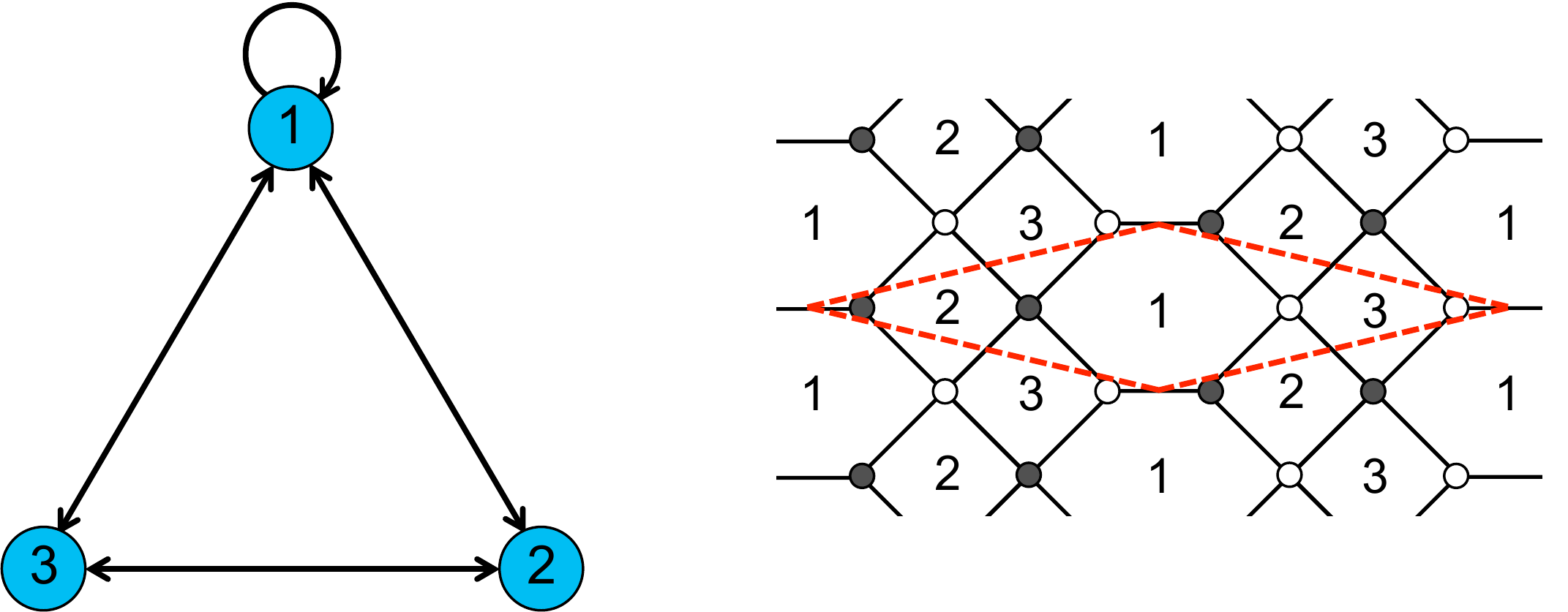}
\caption{Quiver and dimer diagrams for the SPP theory.}
\label{quiver_SPP}
\end{center}
\end{figure}

The web diagram for the SPP singularity and its complex deformation are  shown in \fref{web_SPP}.
The corresponding quiver and dimer diagrams are depicted in Figure \ref{quiver_SPP}.

\begin{figure}[!ht]
\begin{center}
\includegraphics[width=10cm]{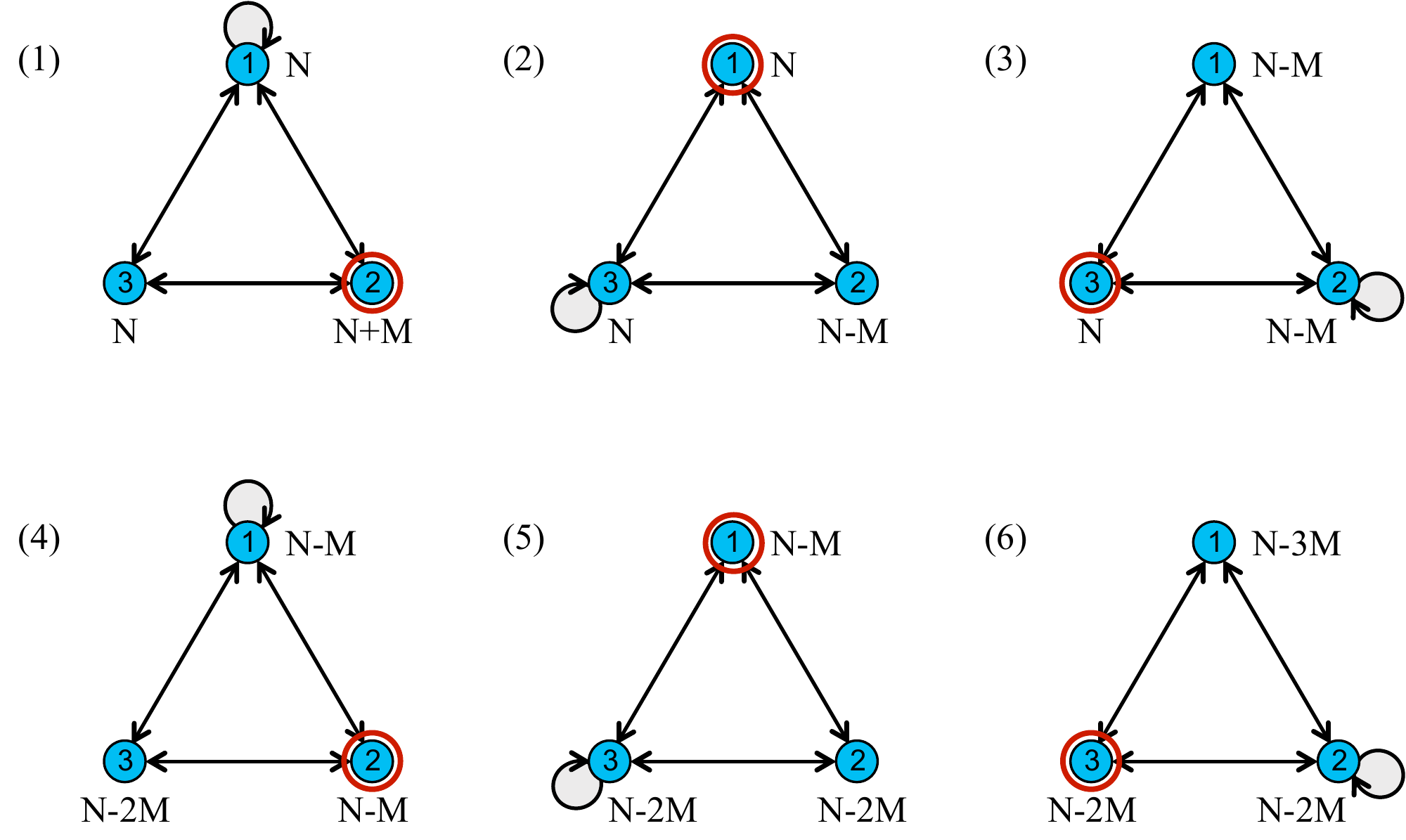}
\caption{Sequence of quivers in one period of the SPP cascade. We have indicated in blue the dualized node at each step.}
\label{quivers_cascade_SPP}
\end{center}
\end{figure}

The theory is non-chiral, so the ranks of the gauge factors are arbitrary, leading to two independent kinds of fractional branes. We consider the following choice or regular and fractional branes, which leads to a duality cascade and the IR complex deformation in \fref{web_SPP}(b)
\beqa
N_1=N_3=N\quad , \quad N_2=N+M \ .
\eeqa
The cascade proceeds with a period given by the consecutive dualizations in a sequence $(2,1,3,2,1,3)$, as shown in \fref{quivers_cascade_SPP}. In each period, the number of regular D3-branes decreases as $N \rightarrow N-3M$.

\begin{figure}[!ht]
\begin{center}
\includegraphics[width=13cm]{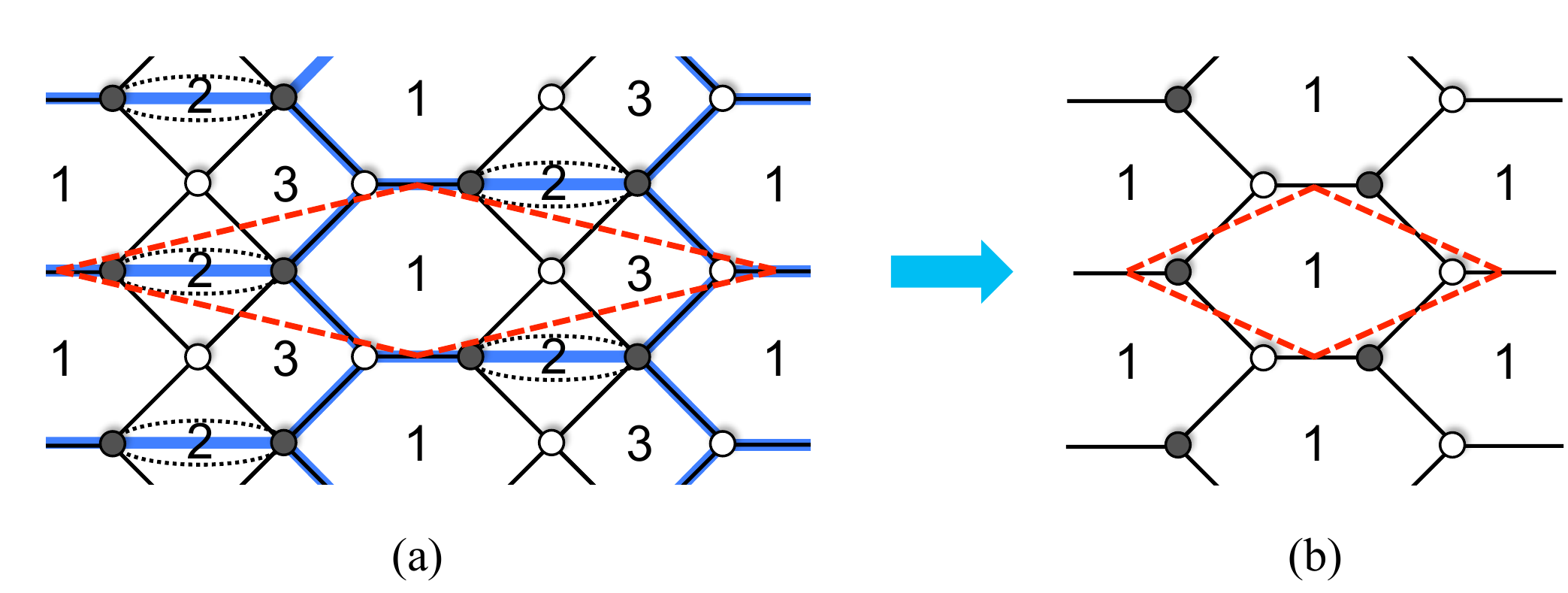}
\caption{Strong dynamics of the SPP theory (a), triggering a complex deformation to the $\mathbb{C}^3$ theory (b).}
\label{dimer-deformation-spp}
\end{center}
\end{figure}

At the IR end of the cascade (and for suitable choice of the UV value of $N$) the theory runs out of D3-branes and the strong dynamics of the gauge factor 2 triggers a complex deformation of the geometry. Its effect on the gauge theory can be easily displayed on the dimer diagram, see \fref{dimer-deformation-spp}. This results in the dimer of $\mathbb{C}^3$, showing that the complex deformation smoothes out the singularity completely.
%

\medskip

\begin{figure}[!ht]
\begin{center}
\includegraphics[width=10cm]{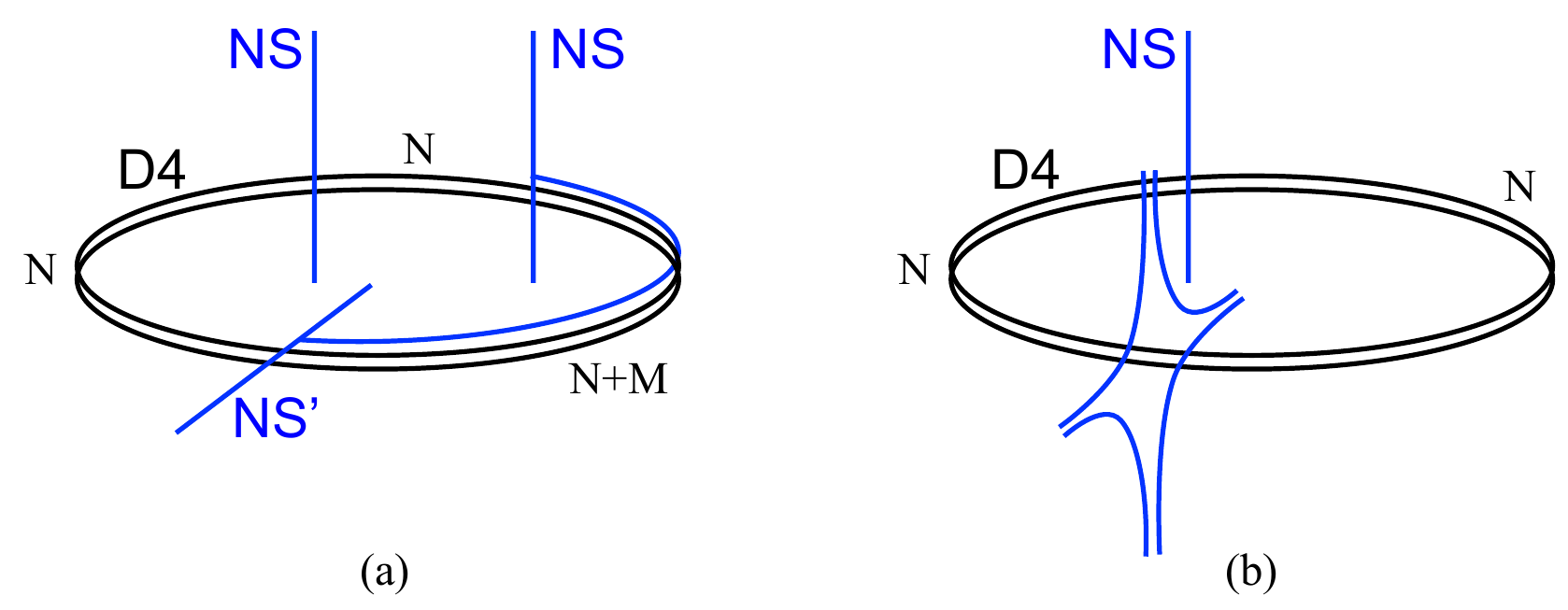}
\caption{(a) Hanany-Witten configuration of NS5-branes and D4-branes, providing the T-dual to D3-branes at the SPP singularity. Figure (b) describes the T-dual of the complex deformed geometry by the recombination of relatively rotated NS5-branes.}
\label{spp-hw}
\end{center}
\end{figure}

The theory of D3-branes at an SPP singularity admits a T-dual description \cite{Uranga:1998vf} in terms of a Hanany-Witten configurations \cite{Hanany:1996ie} of D4-branes stretched between NS5-branes, see \fref{spp-hw}(a). In this picture, the duality cascade corresponds to the motion of the NS'-brane around the $\IS^1$, with crossings with the NS-brane describing Seiberg dualities, which change the number of D4-branes by (the reverse) brane creation effect. The IR deformation corresponds to the recombination of the NS'-brane and one of the NS-branes when the number of regular D4-branes is exhausted, see \fref{spp-hw}(b).

\bigskip

\subsection{Adding Orbifolds}

The $\mathbb{Z}_3$ orbifold in the main text belongs to a general class of orbifolds of SPP.
It is possible to quotient the HW picture of the SPP  by a $\mathbb{Z}_k$ action
\beqa
z\to e^{\frac{2\pi i}{k}} z\quad ,\quad w\to e^{-\frac{2\pi i}{k}} w \ , 
\eeqa
where $z$, $w$ denote the coordinates along the NS-, NS'-branes respectively. This class of orbifolds (originally proposed in \cite{Lykken:1997gy}), has been recently studied in \cite{Franco:2015jna} in the context of class $S_k$ theories \cite{Gaiotto:2015usa}, in setups with non-compact $x^6$.

These theories still correspond to toric geometries, and have a simple dimer diagram. Following the recipe in \cite{Franco:2005rj}, one simply increases (by a factor $k$) the basic unit cell of the parent theory. Therefore each ingredient of the original dimer (face, edge or vertex) has $N$ copies under the quantum $\mathbb{Z}_k$ symmetry of the orbifold. The orbifold theories inherit certain phenomena from the parent theory, like the duality cascades or the IR complex deformations, as described in \sref{sec:parent-spp-z3}.

\bigskip

\subsection{SPP with an Orientifold Line}
\label{app:spp-line}

\begin{figure}[!ht]
\begin{center}
\includegraphics[width=7cm]{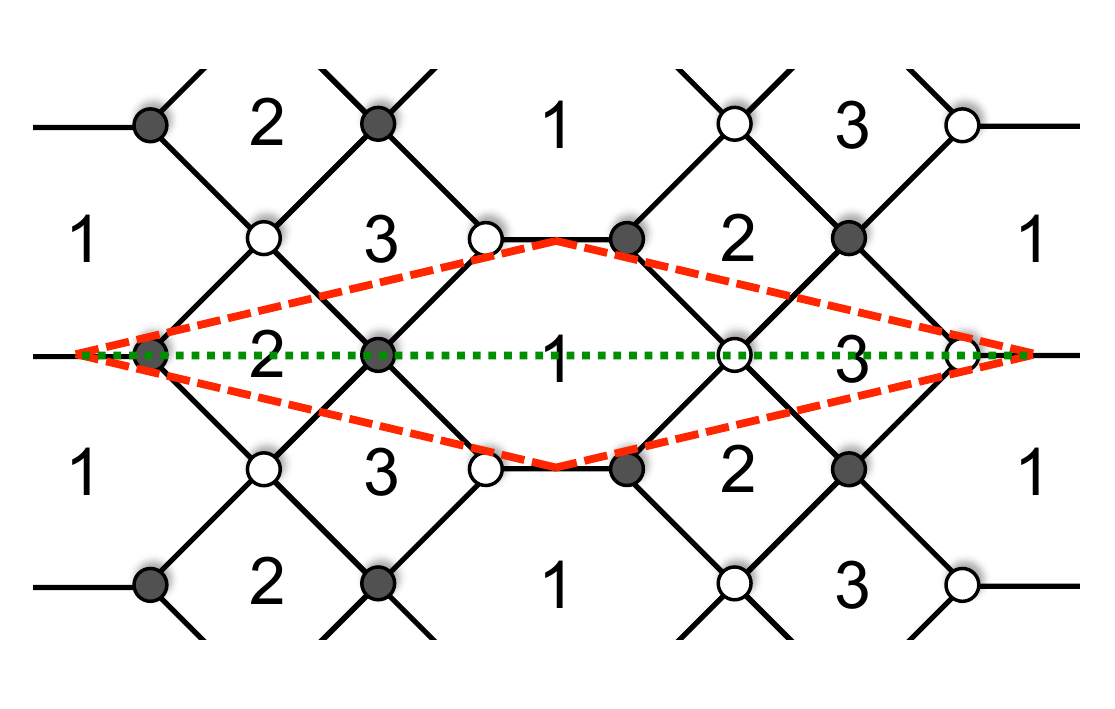}
\caption{Dimer diagram for SPP with a line orientifold.}
\label{spp-dimer-oline}
\end{center}
\end{figure}

In this section we discuss the SPP in the presence of an extra orientifold projection. Orientifolds of toric singularities can be systematically classified and studied using the techniques in \cite{Franco:2007ii}, based on symmetries of the dimer diagrams. In this section we focus on an orientifold with a fixed line, shown in Figure \ref{spp-dimer-oline}, which is related to the model of section \ref{sec:unorient-spp-cascade} before orbifolding. As in there, we choose the orientifold sign projecting invariant gauge factors onto symplectic ones, and invariant edges onto two-index antisymmetric representations.

The model has a T-dual Hanany-Witten construction, corresponding to the addition of an O8-plane stretching in the 4d spacetime, times the $\IS^1$ direction times the $z$ and $w$ complex planes \cite{Feng:2001rh}. Therefore each NS- or NS'-brane is mapped to itself, and so is each D4-brane interval, i.e. each gauge factor. This agrees, with the dimer in Figure \ref{spp-dimer-oline}. 
In the resulting theory each face describes a symplectic factor, the bifundamentals $X_{ij}$ and $X_{ji}$ are exchanged, so one linear combination survives, and the adjoint chiral multiplet $X_{11}$ is projected down to the antisymmetric representation. 

We can also introduced flavor from D7-branes in the original theory, which in the HW T-dual correspond to D8-branes on top of the O8-plane. In the dimer diagram they are described following \cite{Franco:2013ana} (see also \cite{Franco:2006es}).

\begin{figure}[!ht]
\begin{center}
\includegraphics[width=13cm]{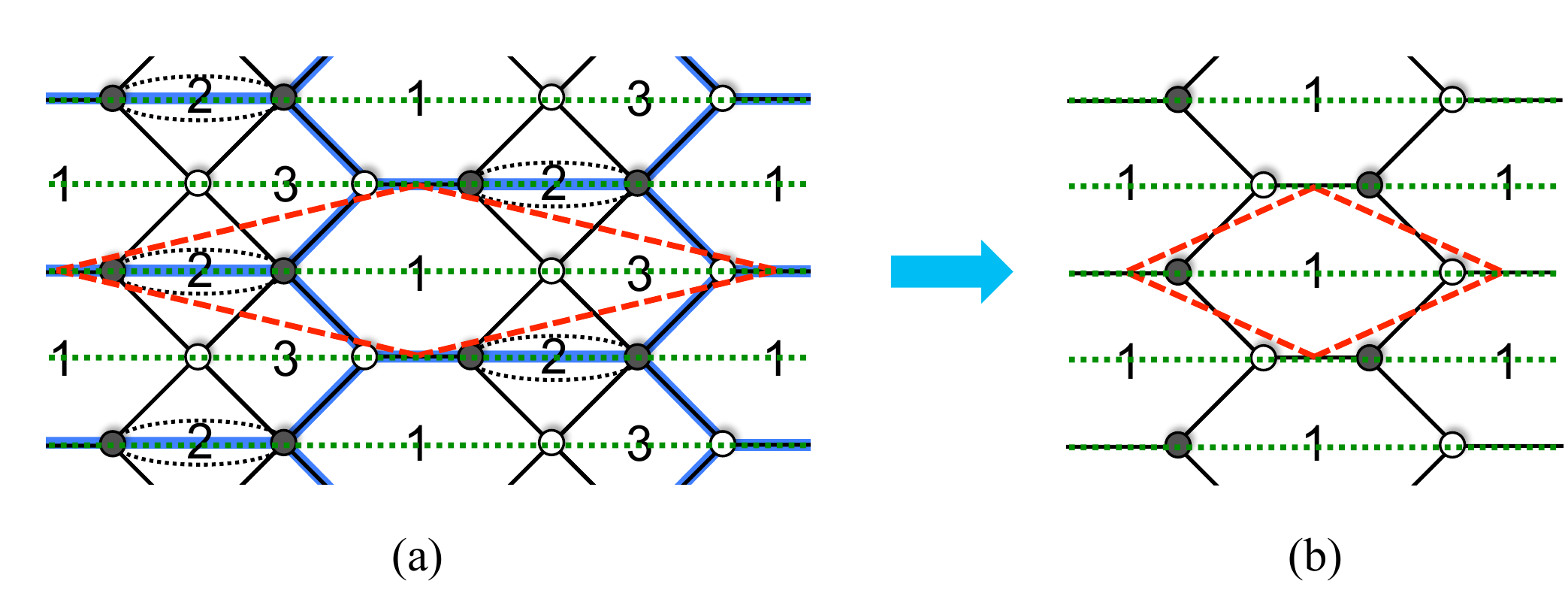}
\caption{Complex deformation of the SPP with a line orientifold.}
\label{spp-oline-deformed}
\end{center}
\end{figure}

As is clear from the dimer, the fractional branes triggering the cascade and complex deformation are invariant under the orientifold action. Hence the orientifold theory has a duality cascade and a complex deformation to an orientifold of $\mathbb{C}^3$. This is also manifest in the HW T-dual, since the recombination of the NS- and NS'-branes into a smooth curve $zw=\epsilon$ is compatible with the O8-plane projection (which leaves $z$ and $w$ invariant. The description of the complex deformation in the field theory can be carried out in detail, and amount to the dimer diagram operation shown in \fref{spp-oline-deformed}.  In other words, the deformation proceeds as a $\IZ_2$ invariant version of the deformation in the parent oriented SPP theory.


\bibliographystyle{JHEP}
\bibliography{mybib}


\end{document}

%% file: pref.tex
\newcommand{\be}{\begin{equation}}
\newcommand{\ee}{\end{equation}}
\newcommand{\beq}{\begin{equation}}
\newcommand{\beql}[1]{\begin{equation}\label{#1}}
\newcommand{\eeq}{\end{equation}}
\newcommand{\ba}{\begin{array}}
\newcommand{\ea}{\end{array}}
\newcommand{\bea}{\begin{eqnarray}}
\newcommand{\beal}[1]{\begin{eqnarray}\label{#1}}
\newcommand{\eea}{\end{eqnarray}}
\newcommand{\ben}{\begin{enumerate}}
\newcommand{\een}{\end{enumerate}}
\newcommand{\bean}{\begin{eqnarray*}}
\newcommand{\eean}{\end{eqnarray*}}
\newcommand{\eref}[1]{(\ref{#1})}
\newcommand{\sref}[1]{\S\ref{#1}}

\newcommand{\fref}[1]{Figure \ref{#1}}
\newcommand{\btab}[1]{\begin{tabular}{#1}}
\newcommand{\etab}{\end{tabular}}

\newcommand{\comment}[1]{}

\newcommand{\IC}{\mathbb{C}}

\newcommand{\qed}{\nobreak \ifvmode \relax \else
      \ifdim\lastskip<1.5em \hskip-\lastskip
      \hskip1.5em plus0em minus0.5em \fi \nobreak
      \vrule height0.75em width0.5em depth0.25em\fi}